%% file: main.tex
\documentclass{sn-jnl}
\input{sections/header}

\input{sections/authors}
\date{\today}

\begin{document}
\title{Crypto-asset Taxonomy for Investors and Regulators}
\input{abstract}
\maketitle

\input{sections/introduction}

\input{sections/related_works}

\input{sections/methodology}

\input{sections/taxonomy_dimensions}

\input{sections/dataclass}

\input{sections/case_studies}

\input{sections/conclusion_and_discussion}

\FloatBarrier
\clearpage
\nocite{*}
\bibliographystyle{plain}
\bibliography{references}

\input{sections/appendix}

\end{document}

%% file: sections/header.tex
\usepackage{enumitem}
\usepackage{natbib} 
\usepackage{graphicx}
\usepackage{xcolor}
\usepackage{graphicx}
\usepackage{subfig}
\usepackage{placeins}
\usepackage{minted}
\usepackage[T1]{fontenc}  
\usepackage[utf8]{inputenc} 
\usepackage{lmodern}
\usepackage[most]{tcolorbox}
\usepackage{booktabs}
\usepackage{tabularx}
\usepackage{listings}
\usepackage{acro}
\usepackage{csquotes}
\usepackage{longtable}
\usepackage{ragged2e}

\newcolumntype{L}[1]{>{\RaggedRight\arraybackslash}p{#1}}

\lstdefinelanguage{json}{
  basicstyle=\ttfamily\small,
  showstringspaces=false,
  morestring=[b]",
  morecomment=[l]{//},
  commentstyle=\color{gray},
  stringstyle=\color{black},
}
\newtcolorbox{CaseBox}[1][]{
  breakable, colback=white, colframe=black!24, title=#1,
  fonttitle=\bfseries, boxsep=1ex, arc=2mm, left=1ex,right=1ex,top=1ex,bottom=1ex
}

\acsetup{
  first-style = long-short, 
}

\DeclareAcronym{MiCAR}{
  short = MiCAR,
  long  = Markets in Crypto-Assets Regulation
}

\DeclareAcronym{mifid}{
  short = MiFID,
  long  = Markets in Financial Instruments Directive
}

\DeclareAcronym{esma}{
  short = ESMA,
  long  = European Securities and Markets Authority
}

\DeclareAcronym{dlt}{
  short = DLT,
  long  = distributed ledger technology
}

\DeclareAcronym{defi}{
  short = DeFi,
  long  = decentralised finance
}

\DeclareAcronym{cefi}{
  short = CeFi,
  long  = centralised finance
}

\DeclareAcronym{tradfi}{
    short = TradFi,
    long = traditional finance
}

\DeclareAcronym{dao}{
  short = DAO,
  long  = decentralised autonomous organisation
}

\DeclareAcronym{nft}{
  short = NFT,
  long  = non-fungible token
}

\DeclareAcronym{tvl}{
  short = TVL,
  long  = total value locked
}

\DeclareAcronym{psm}{
  short = PSM,
  long  = Peg Stability Module,
  short-plural = PSMs,
  long-plural = Peg Stability Modules
}

\DeclareAcronym{emt}{
  short = EMT,
  long = e-money token
}

\DeclareAcronym{art}{
short = ART,
long = asset-referenced token,
short-plural = ARTs,
long-plural = asset-referenced tokens
}

\DeclareAcronym{lst}{
  short = LST,
  long  = liquid staking token,
  short-plural = LSTs,
  long-plural = liquid staking tokens
}

\DeclareAcronym{ptc}{
  short = PTC,
  long =  pass-through certificate
}

\DeclareAcronym{pos}{
  short = PoS,
  long  = proof of stake
}

\DeclareAcronym{pow}{
  short = PoW,
  long  = proof of work
}

\DeclareAcronym{unl}{
  short = UNL,
  long  = Unique Node List
}

\DeclareAcronym{mdt}{
short = MDT,
long = Minimum Decentralisation Test
}

%% file: sections/authors.tex
\author[1]{Xiao Zhang}
\author[1,2]{Juan Ignacio Iba\~nez}
\author[1,2]{Jiahua Xu}

\affil[1]{Exponential Science}
\affil[2]{Department of Computer Science, University College London}

%% file: abstract.tex
\abstract{ Crypto-assets are a main segment of electronic markets, with growing trade volume and market share, yet there's no unified and comprehensive asset level taxonomy framework. This paper develops a multidimensional taxonomy for crypto-assets that connects technical design to market structure and regulation. Building on established taxonomy guideline and existing models, we derive dimensions from theory, regulatory frameworks, and case studies. We then map top 100 assets within the structure and provide several detailed case studies. The taxonomy covers technology standard, centralisation of critical resources, asset function, legal classification and mechanism designs of minting, yield, redemption. The asset mapping and case studies reveal recurring design patterns, capture features of edge cases that sit on boundaries of current categorisations, and document centralised control of nominal decentralised assets. This paper provides framework for systematic study for crypto markets, supports regulators in assessing token risks, and offers investors and digital platform designers a tool to compare assets when building or participate in electronic markets.
} 
\keywords{Crypto-assets, MiCAR, Regulatory classification, Token design, Decentralized finance, Liquid staking tokens(LST)}

%% file: sections/introduction.tex
\section{Introduction and Motivation}
Digital assets are assets issued and transferred through distributed ledger or blockchain technology\cite{ U.S.SecuritiesandExchangeCommissionSEC2019FrameworkAssets}. They emerged with Bitcoin, which combined pre-existing building blocks such as distributed ledgers, public-key cryptography and consensus protocols to support a peer-to-peer electronic cash system, and have since evolved into a wide family of blockchain-based architectures. This broad category includes both fungible and non-fungible instruments, ranging from cryptocurrencies and stablecoins to \ac{nft}s. 
Following MiCAR and IOSCO, we define crypto-assets as fungible tokens recorded on distributed ledgers within digital-asset class, encompassing stablecoins, governance tokens, and other transferable instruments.  Our taxonomy focuses on this subset, as it underpins most of decentralised assets. 

Despite the rapid growth of crypto-assets, there is no unified taxonomy that consistently categorizes assets in a way that is clear, non-overlapping, and useful for regulators and investors. From regulating and investing views, existing taxonomies exhibit important shortcomings. Some remain abstract and detached from economic or regulatory realities (e.g.\ The Token Taxonomy Framework (TTF) in \autoref{fig:tax-ttf-full}, \cite{InterWorkAlliance2020InterWorkFinal}). Some are ambiguous and overlapping (e.g., BIS taxonomy in \autoref{fig:tax-bis-full} , \cite{Auer2023TheDeFi}). Some are essentially market-driven and omit regulatory and functional considerations (e.g.\ MarketVector’s taxonomy in \autoref{fig:tax-mv-full} , \cite{Leinweber2024AWorld}). Also, emerging crypto-asset categories such as \ac{lst}s challenge existing taxonomies. The systematisation in \cite{Gogol2024SoK:Risks} demonstrates that LSTs introduce entirely new design dimensions, validator-selection models, distributed validator technology (DVT), reward-distribution structures, and pegging mechanics. Regulatory analyses in \cite{HoganLovells2025RegulatingGuide} highlight that legal systems lack a coherent categorization for \ac{lst}s and that \ac{lst}s blur boundaries between staking, custody, investment schemes, and lending. \cite{Scharnowski2025TheDiscovery} provides a broader economic view emphasizing \ac{lst}s simultaneously function as collateral, yield-bearing instruments, and transferable assets. 

While previous taxonomies mainly focus on classifications on systems and protocols, we aim to build taxonomy that aligns technical, mechanistic and regulatory dimensions on crypto-asset level. Our contribution lies in three folds: the taxonomy framework captures the design processes underlying crypto-assets with an emphasis on decentralised assets, provides a comprehensive, rigorous, and consistent foundation that remains scalable as new asset classes and protocols develop; furthermore, within the proposed taxonomy we identify \ac{tradfi} analogues for assets that exhibit specific combinations of features, these analogies help situate crypto-assets within familiar economic and regulatory categories in \ac{tradfi}, thereby providing clearer guidance for both regulators and investors on classification and risk assessment; last, we empirically documented top 100 assets, illustrating how assets could be mapped into this taxonomy, revealing asset design patterns and creating a reusable dataset.

%% file: sections/related_works.tex
\section{Related Work}
Clear taxonomies are essential for coherent oversight, \cite{Zetzsche2024CryptoCustody} show that under MiCA, asset categorisation directly determines custody obligations and reveals gaps between \enquote{institutional} and \enquote{asset} resilience.\cite{MilkenInstitute2023AAssets} highlights fragmented regulatory definitions and argues that consistent asset categorisation is a prerequisite for supervision. 
\cite{Breydo2024TheElusive} shows how mislabelled \enquote{utility tokens} and unclear economic rights obstruct enforcement and consumer protection, advocating an asset-level taxonomy grounded in substantive rights. Following same motivation, \cite{Schuler2024OnFinance} introduce a decentralisation-vector taxonomy distinguishing genuine \ac{defi} from \enquote{on-chain \ac{cefi}}, helping regulators identify control points and apply \enquote{same risk, same rules}.
\cite{Yuyama2024ProposalRegulation} argues that meaningful oversight of DeFi is impossible without a clear regulatory framework grounded in taxonomy. The authors show that regulators can target disclosure and accountability requirements by where the control exists, governance keys, upgrade rights, incentives, or data flows.

Prior taxonomies mainly focus on blockchain systems, applications, or business models, rather than individual crypto-assets. \cite{Tasca2019AClassification} decomposed systems into consensus, data, transaction and network layers, yielding a structured tree useful for technical comparisons but unable to capture the post-2020 proliferation of assets and protocols. The BIS DeFi Stack Reference (DSR) model \cite{Auer2023TheDeFi} extends this view with a layered framework (settlement, application, interface) that distinguishes assets, protocol types, and compositions, while noting the rise of wrapped assets, governance tokens, and associated centralisation risks. \cite{Sai2021TaxonomyReview} built a systematic taxonomy of centralisation risks across architectural layers, showing how power can concentrate in governance, mining, software clients, and network structure. Application-level surveys like \cite{Abdelmaboud2022BlockchainDirections} organise blockchain-for-IoT solutions by blockchain type, consensus, smart-contract use, and domain, focusing on security and trust properties.

Beyond general frameworks, SoK studies and surveys extend DeFi classification by categorizing protocols and risks \cite{Werner2022SoK:DeFi}, proposing entropy-based measures of decentralisation across consensus, network, governance, wealth, and transactions \cite{Zhang2023SoK:Decentralization}. \cite{Gogol2024SoK:Risks} provide a data-driven taxonomy of DeFi protocols covering over 85\% of TVL, classifying them into liquidity pool protocols, synthetic token protocols, and aggregators. By mapping a wider set of newer protocols than earlier studies, their framework links protocol risk to three dimensions: protocol design, stakeholder type, and underlying token, offering one of the most comprehensive taxonomic overviews of the \ac{defi} landscape. 
Taxonomies that classify segments of block systems also help illuminate how to structure a coherent, asset-level taxonomy. \cite{Moin2019ADesigns} propose a design-space taxonomy for stablecoins along four main axes: peg choice, collateral type, collateralisation level, and \enquote{mechanism of action}. \cite{Cousaert2022SoK:DeFi} systematise DeFi yield at the protocol level by distinguishing three fundamental yield sources, \cite{Ito2024CryptoeconomicsOpinions} argues framing yield and token rewards as part of a general mechanism-design problem.

In summary, while prior taxonomies and surveys provide valuable insights into either technical architectures, regulatory categories, or financial protocol families, they remain fragmented and often omit the interplay between asset design, cross-chain deployment, financial utility, and structural risk. 
Our contribution is to advance a unified taxonomy that combines functional, structural, technical, financial dimensions, offering a more comprehensive descriptive map and a practical basis for risk analysis and regulatory dialogue.

%% file: sections/methodology.tex
\section{Methodology}
Our taxonomy is designed to organise the crypto-assets at the token level. We focus mainly on fungible tokens that could be transferred and held as assets on balance sheet. We intend to develop descriptive feature dimensions without prescribing how specific assets should be regulated or designed. In order to construct a multi-dimensional taxonomy, this paper employs a three-step methodology. 

First, we conduct a structured review of existing taxonomies of blockchain and crypto-asset technologies, as well as surveys that focus on specific segments of the crypto ecosystem. Also, to ensure compatibility with existing classifications, we incorporate current legal frameworks for crypto-asset regulation from EU and US, such as \ac{MiCAR}, \ac{mifid}, and the Howey Test \cite{EuropeanParliament2023Regulation2019/1937}\cite{EuropeanParliament2014Directive2011/61/EU}\cite{SupremeCourtoftheUnitedStates1946SECCo.}. From this literature and related regulatory and industry reports, we extract the key dimensions and candidate categories that are relevant at the asset level. We iteratively refine these candidate dimensions by merging overlapping concepts and eliminating redundancies, aiming for orthogonal, non-overlapping categories. Ambiguous cases in the literature were used to stress-test the dimensions and clarify decision rules. These insights are consolidated into a unified taxonomy framework with orthogonal, non-overlapping dimensions to ensure each aspect of a crypto-asset is captured distinctly.

Then, in line with established design principles for taxonomy development, we examine representative assets spanning major functional types (e.g., native layer-1 tokens, wrapped assets, \ac{lst}s). For each representative asset, we conduct more detailed case studies structured by the proposed taxonomy’s dimensions, mapping the asset’s features to the appropriate categories. This step helps illustrate and refine the taxonomy. Where applicable, we identify \ac{tradfi} analogies by mapping the observed features to parallel categorisations in existing financial instruments and regulatory classifications, highlighting how similar economic functions are categorised in conventional frameworks. The case studies can be found in \autoref{app: case study}.

Finally, we operationalise the taxonomy in code by implementing a CryptoAsset dataclass that encodes the relevant attributes for each dimension as attributes of asset instances. We instantiate top 100 assets by market capitalisation from CoinGecko as of 19 Nov 2025, and by classifying each of these top assets according to our taxonomy, from sources like project documents, protocol whitepapers, and legal texts. We then obtain a set of asset \enquote{buckets} and summary statistics for each category. Then, we summarise how economic functions and technical designs are distributed across the current market. The resulting distribution of assets across categories provides insights into how economic functions and technical designs are represented in today’s crypto market. The design and decision rules of our dataclass can be found in \autoref{sec:dataclass_design}. 

This methodology is necessarily constrained by the availability and quality of project documentation and legal guidance, and some borderline cases require judgement calls. We therefore view the taxonomy as a structured, transparent starting point rather than a definitive classification of all existing and future assets.

%% file: sections/taxonomy_dimensions.tex
\section{Taxonomy Dimensions}

This section focuses on dimensions that matter for regulation and risk, including who issues or governs the asset, how units are created and brought into circulation, what rights, cash flows, or claims users have, and how redemption, custody, and control over the backing resources are structured. We adopt this regulatory lens to make risk pathways and compliance obligations comparable across assets. Descriptive market metrics such as prices, volatility, trading volumes, \ac{tvl} or market share and size-based rankings are excluded from the taxonomy.

\subsection{Technical Standard}
Crypto-assets follow different technical templates that define how they are issued and transferred across blockchains. The choice of standard is not only a technical detail 
but shapes how token economies evolve: fungible standards such as ERC-20 create large, exchange-driven networks, while non-fungible standards such as ERC-721 lead to more decentralised, peer-to-peer transfer patterns \cite{Loporchio2024ComparingNetworks}. 
Because these structural differences yield distinct economic behaviours, they should be distinguished for regulatory categorisation.

\begin{itemize}[noitemsep]
    \item Native: Assets built on their own chain and governed by its own protocol rules rather than a smart-contract token standard. (e.g., BTC, ETH, TON).
    \item ERC-20: Fungible, widely used standard (e.g., WETH, MKR). Its fungibility enables broad DeFi integration and indicates wider contagion surface during stress and protocol dependencies.
    \item Other token templates across blockchains (e.g. AVAX on ERC-721 standard, BNP Paribas on ERC-1400 / ERC-3643 standards).
\end{itemize}

Although the technical standard by itself does not determine legal classification, regulators globally assert that classification must be based on the asset's economic function and potential rights attached to such asset, if any. Leading regulatory bodies, such as the ESMA, reinforce this by emphasizing the principle of technological neutrality: the classification of crypto-assets must be based on their economic function and the rights they confer, rather than their underlying technology\footnote{See in \cite{ESMA2024ESMA75453128700-1323Instruments}, ESMA Final Report, Guidelines on the conditions and criteria for the qualification of crypto-assets as financial instruments under \ac{mifid} II, ESMA75-453128700-1323, 17 Dec. 2024, para. 56.}. Nonetheless, the technical standard remains an important consideration that, in interplay with other categories, contributes to the overall assessment of a crypto-asset. 


\subsection{Centralisation Dimension}
\label{sec: centralisation_dimension}
A fundamental distinction in any taxonomy of crypto-assets concerns the degree of centralisation in governance and custody.  This distinction is not only emphasised in the academic literature \cite{Schuler2024OnFinance}, but also explicitly recognised by regulators. EU \ac{MiCAR} define crypto-assets with no identifiable issuer as decentralised \footnote{MiCAR Regulation (EU) 2023/1114: \enquote{Where crypto-assets have no identifiable issuer, they should not fall within the scope of Title II, III or IV of this Regulation}.} and define crypto-asset services as decentralised when there's no intermediary\footnote{MiCAR  Regulation (EU) 2023/1114: \enquote{where crypto-asset services are provided in a fully decentralised manner without any intermediary, they should not fall within the scope of this Regulation}}. The recent US federal bill on digital asset market structure, Digital Asset Market Clarity Act of 2025 (the CLARITY Act), proposes a rule-based framework and quantitative threshold of 20\% for both governance power and token supply.

There are existing measures for centralisation and decentralisation. Decentralisation is measured across five facets: architectural, political, logical, geographical, and resource as identified in \cite{Zhang2023SoK:Decentralization}, they applied standard concentration measures, such as Shannon entropy, Gini, HHI and the Nakamoto coefficient on each facet measuring centralisation. Similarly, \cite{Ovezik2025SoK:Decentralization} stratify decentralisation into eight layers (hardware, software, network, consensus, tokenomics, client API, governance, and geography) and introduce the \ac{mdt}. Under this criterion, a system fails the MDT if there exists any layer in which a single legal entity can control enough of the relevant parties governing a critical resource to endanger core system properties such as safety, liveness, privacy, or stability.

Following our taxonomy structure and incorporating some indicators from \cite{BlackVogel2025DARTEWashington}, we apply the \ac{mdt} method introduced in \cite{Ovezik2024SoK:Decentralization}, and assess centralisation across 11 functional sub-dimensions: governance rule change authority, governance voting power, minting authority, minting data and parameter control, yield reward policy control, yield operator and distribution control, redemption reserve control, and redemption mechanism control. For each sub-dimension, we identify the critical resources. See all critical resources listed in \autoref{tab: critical_resources}. The centralisation categorisation of each domain is then determined by number of entities in control of all resources under this domain. Applying the \ac{mdt}, a domain is deemed decentralised if no single legal entity can unilaterally compromise the resource’s operation or alter its outcomes. We further aggregate these sub-domains into six functional groups: governance (rule changes, voting), minting (mint authority, mint data and parameters), yield (reward policy, yield distribution), redemption (reserves, redemption mechanism), market (ownership makeup, exchange control), community (community transparency). We ignore groups that are not applicable (e.g., yield for non-yielding assets, redemption for non-referenced assets); then, an asset is classified as decentralised if all groups pass the MDT, centralised if the first 4 groups are subject to unilateral control by a single entity, and hybrid otherwise, reflecting partial decentralisation across control surfaces. In other words, we treat decentralisation as a non-compensatory property: centralised control over any core critical resource (governance, minting, yield, redemption) is sufficient to classify an asset as centralised at the core. The rest two dimensions of market and community can only add further evidence of centralisation, but a decentralised trading environment does not overturn centralised control at the core.

\begin{table}[ht]
\centering
\caption{Critical Resources}
\begin{tabular}{L{2.2cm} L{4.8cm} L{5.2cm}}
\toprule
\textbf{Subdimensions} & \textbf{Description} & \textbf{Resources} \\ 

Rule change authority 
& who can propose, modify, or execute protocol-level rule changes 
& admin keys; upgrade authorities; emergency pause/shutdown \\ 

Voting power 
& how decision-making power is distributed and exercised 
& quorum/threshold; eligible voters; delegation rules; validator curation \\

Mint authority 
& who is technically or legally able to mint or burn tokens 
& mint keys; whitelisted minters; custodial issuers \\ 

Data and parameter control 
& who controls the data and risk parameters that govern issuance 
& oracle operators; oracle aggregators; collateral parameters; liquidation rules \\ 

Reward policy control 
& who determines the level and timing of rewards 
& reward rates; emission schedules; fee levels \\ 

Operator and distribution control 
& who manages operational entities and routes yield flows 
& validator/operator selection; stake allocation; distribution mechanisms \\ 

Reserve control 
& how backing assets are managed 
& reserve custodians; reserve composition; attestations/audits \\ 

Mechanism control 
& who controls user access to redemption and the mechanics of redeeming 
& gatekeepers/whitelisting; redemption queues; settlement custodians; freeze/blocklist powers \\ 

Ownership makeup
& who owns big chunks of the supply
& on-chain holder; off-chain register; distribution disclose \\ 

Exchange control
& who controls where and how it trades
& exchange listing; freeze/halt controls; designated market makers \\ 

Community transparency
& who can see the system changes and management
& project info; governance info; operational info \\
\bottomrule
\end{tabular}
\label{tab: critical_resources}
\end{table}

\begin{itemize}
    \item Centralised: The asset’s governance, minting or redemption are controlled by a single entity or tightly-layered group; token-holder influence is limited; issuer liability exists. Such as custodial stablecoins (e.g., USDT, USDC), exchange tokens (e.g., BNB, FTT), \ac{lst}s where redemption and operations are managed by a central entity (e.g., cbETH, BETH). Custodial lending platforms such as Celsius and BlockFi. 
    \item Decentralised: Governance, issuance, redemption and yield mechanisms are broadly distributed; no single entity exercises unilateral control; token-holders or protocol participants hold meaningful influence. Such as decentralised stablecoins (e.g., DAI), governance tokens (e.g., UNI, MKR, COMP), \ac{lst}s with decentralised validator sets (e.g., stETH, rETH). Non-custodial DeFi protocols like Aave and Uniswap.
    \item Hybrid: Mixed or transitional models: some domains are decentralised, others remain centrally controlled; governance or redemption may gradually shift over time. Such as a algorithmic stablecoins (e.g., UST), governance tokens of protocols with residual centralisation (e.g., LDO), \ac{lst}s with partial validator centralisation (e.g., ankrETH).
\end{itemize}

The degree of centralisation significantly influences which entity may be designated as an issuer under MiCAR \footnote{See Art. 3(1))}, and consequently whether liability and disclosure obligations attach. A token classified as centralised, where governance, issuance, yield and redemption are controlled by a single legal entity, is far more likely to trigger issuer regimes (such as a required white-paper or authorisation) because the controlling entity can satisfy issuer duties under MiCAR. In contrast, a decentralised token, where the control surface is dispersed and no single entity can unilaterally determine outcomes, tends to shift regulatory scrutiny onto service providers (e.g., CASPs) rather than the issuer. 

\subsection{Asset Function}
Assets can be distinguished by their economic function. While many governance tokens serve only protocol voting functions, some that embed economic or ownership rights may fall under \ac{mifid} II as financial instruments and thus be treated as security tokens rather than within MiCAR\footnote{MiCAR: \enquote{This Regulation does not apply to crypto-assets that qualify as financial instruments as defined in Directive 2014/65/EU (MiFID II).}} \footnote{MiCAR: \enquote{Tokenised financial instruments should continue to be considered as financial instruments for all regulatory purposes.}}.

\begin{itemize}[noitemsep]
    \item Governance tokens: Assets whose defining feature is to convey on-chain governance or control rights over a protocol or pool, such as voting on parameter changes, upgrades, or treasury allocation (e.g., UNI, MKR).
    \item Utility tokens: Following the definition of \ac{MiCAR}\footnote{MiCAR Article~3(1)(9): Utility token is a crypto-asset intended only to provide digital access to a good or service available on \ac{dlt} and accepted solely by its issuer.}, assets whose primary purpose is to provide digital access to a specific application, good or service on a \ac{dlt} system, without granting a claim on external reserves or protocol cash flows, and accepted solely by its issuer. (e.g., SAND, CRO)
    \item Security tokens: Assets whose core economic function is to provide exposure to underlying assets or cash flows, and that therefore behave similarly to traditional securities, fund units, or deposit-like instruments.
\end{itemize}

\subsection{Mechanism Design}
This dimension concerns the design for supply and cash flow along the lifecycle of crypto-assets. We identify three main steps here: (i) Minting, how units are created. Assets are categorised by issuer and the minting mechanism. (ii) Yield, what value added to holders, it has two facets: yield source and distribution mechanism. (iii) Redemption, how assets are linked to other assets. Reference asset and redemption assets are considered. These three subdimensions together shape mechanism design of assets \cite{Auer2023TheDeFi}\cite{Gogol2024SoK:Risks}.

\subsubsection{Minting Type}
A key structural dimension concerns who brings units into existence and how they are created and released into circulation. Minting arrangements combine an institutional component (the presence or absence of an identifiable issuer) with a mechanical component (the rules that determine when new units are minted, against which inputs, and under what constraints).

\paragraph{Issuer}
This dimension captures who (if anyone) creates and makes commitments regarding the token, which is critical for determining the applicability of regulatory regimes such as \ac{MiCAR} and \ac{mifid} II.

\begin{itemize}
    \item Centralised issuer: Assets minted by a clearly identifiable legal person or entity creates, controls issuance, and may promise returns, redemption, or other performance. (e.g., USDC) The unilateral control of issuing may put a person or entity in the position to qualify as an issuer under \ac{MiCAR} Art. 3(1)(19) and Recital (20). In such cases, issuer obligations, such as white-paper requirements and (where applicable) authorisation, would apply.
    \item Protocol issuance: Assets minted or distributed algorithmically via smart contracts, often governed by a DAO or multisig structure, without a single accountable party post-deployment (e.g., DAI, UNI). Assets minted by protocols may not have an issuer in \ac{MiCAR} sense, but CASPs dealing in the token are regulated.
    \item No issuer: Assets minted from consensus mechanisms, without any centralised or delegated issuance process (e.g., BTC, XMR). May be excluded from issuer regimes under \ac{MiCAR}, also unlikely to apply \ac{mifid} II unless intermediaries wrap or repackage them into financial instruments.
\end{itemize}

\paragraph{Minting Mechanism}
Crypto-assets can be created through a variety of minting processes, which differ in their economic logic and technical implementation. Existing surveys and SoK papers identify the following main categories (\cite{Tasca2019AClassification}, \cite{Gogol2024SoK:Risks}, \cite{Auer2023TheDeFi}, \cite{Giulio2021WrappingTokens}):

\begin{itemize}[noitemsep]
    \item Consensus-based minting: \ac{pow} (e.g., BTC), \ac{pos} (e.g., ETH), Proof-of-Authority (e.g., VET), Proof-of-Capacity/Storage (e.g., XCH), and Proof-of-Burn (e.g., XCP).
    \item Collateralised lock-and-mint: Assets minted by depositing collateral into a contract or custodian (e.g., DAI, sUSD).
    \item Staked tokens: Minted by staking PoS tokens to generate staking rewards (e.g., stETH, JotoSOL, stATOM).
    \item Wrapped tokens: Locking the original asset on its own blockchain and minting a copy on another to enable cross-chain use (e.g., wBTC, WETH, cbETH).
    \item Algorithmic burn-and-mint equilibria: Tokens with elastic supply adjusted to maintain a peg (e.g., UST/LUNA pair, AMPL).
    \item Governance- or emission-based minting: Protocol-engineered emissions as incentives or to bootstrap participation (e.g., UNI, COMP).
    \item Pre-mining / initial allocation: Assets minted at launch and distributed to founders or early token buyers (e.g., XRP, EOS).
    \item NFT minting: Token contracts maintain a registry of unique ownership under non-fungible token standards (e.g., Decentraland LAND, BAYC).
\end{itemize}

The legal significance of this dimension is based on the fact that it defines how a crypto-asset originates, determining who, if anyone, can be considered its issuer and which regulatory regime may apply. Under EU law, different minting mechanisms align with categories under \ac{MiCAR} and \ac{mifid} II \footnote{see MiCAR Articles 3 and 4}. Collateralised lock-and-mint models (e.g., DAI) can resemble debt-like instruments and may qualify as ARTs. Liquid staking tokens (e.g., stETH) often reference staked assets and may fall within \ac{mifid} II (if qualifying as derivative) or MiCAR’s scope where a central intermediary exists, while wrapped tokens (e.g., WBTC) may mirror depositary receipts, introducing custodial and redemption obligations. Conversely, consensus-based assets like BTC or ETH, created through decentralised validation, generally fall outside issuer-based regimes. They may nevertheless qualify as so called \enquote{other tokens} under MiCAR.

Under U.S. law, minting determines the degree of managerial control and investor reliance, key elements of the Howey Test. Tokens with pre-mining or centralised issuance (e.g., XRP) show entrepreneurial effort by identifiable actors, increasing the risk of being deemed securities. By contrast, tokens minted through open consensus or algorithmic processes (e.g., BTC) are usually regarded as decentralised commodities.


\subsubsection{Yield Type}
A key distinction in any taxonomy of crypto-asset concerns the yield. Capital appreciation is an increase in the price or value of assets. Yield encompasses all types of income derived from holding an asset, excluding capital appreciation. We classify an asset with its yield source and distribution mechanism.

\paragraph{Yield Source}
We distinguish yield by underlying economic activity (direct yield) or by protocol-engineered emissions (incentive yield). Recent surveys explicitly make this separation. These surveys note that users earn income \enquote{in forms of transaction fees, interest, or participation rewards} while protocols may also \enquote{distribute newly minted tokens to incentivize users to contribute liquidity} \cite{Xu2023ReapProtocols}. 

In \cite{Cousaert2022SoK:DeFi}, yield source is further categorised to borrowing demand, liquidity mining, and revenue sharing, \cite{Scharnowski2025TheDiscovery} emphasizes that staking yields stem directly from protocol-level block rewards and transaction fees rather than purely inflationary incentives.

Importantly, some assets do not generate yield at all, functioning purely as stores of value or transactional media. The source of the yield determines the core economic activity that generates profit, directly addressing the critical fourth prong of the US Howey Test, profit derived from the entrepreneurial or managerial efforts of others.

\begin{itemize}
    \item Direct Yield: derived from actual economic activity.
    \begin{itemize}
        \item Lending/Borrowing: Interest generated from borrowers in protocols for loanable funds, which accrues to depositors via interest-bearing tokens (e.g., aUSDC, cDAI).
        \item Liquidity Provision Fees: Trading fees rewarded to liquidity providers from automated market maker protocols such as Uniswap and Curve (e.g., ETH/USDC, DAI/USDT, WBTC/ETH pools).
        \item Staking rewards: Validator rewards distributed on \ac{pos} chains, such as Ethereum, Polkadot, and Cosmos (e.g., ETH, DOT, ATOM), and on liquid staking tokens (e.g., mSOL, stETH, BETH). Direct yield from staking may often causes \ac{lst}s to be classified as \ac{art}s because they reference the value of the staked asset plus its generated rewards, which are tied to the economic activity of the underlying chain

        \item Revenue Sharing: Collected revenues redistributed to token stakers(e.g., xSUSHI from SushiSwap, GMX, DYDX). 
    \end{itemize}
    \item Incentive Yield: created by minting new tokens or elastic supply mechanisms. Yield generated by a protocol via new token emissions (e.g., liquidity mining, governance rewards) can be scrutinised as evidence of ongoing managerial efforts by the founding team or promoter to incentivize participation and bootstrap the network. This continuous intervention raises the risk that the expectation of profits is derived from those human efforts, potentially triggering security classification.
    \begin{itemize}
        \item Liquidity Mining Incentives: Minted and distributed governance tokens (e.g., COMP, CRV, SUSHI) to motivate participants from protocols such as Compound, Curve, and SushiSwap.
        \item Burn-and-Mint Equilibria: Value redistributed after supply adjustment in algorithmic stablecoins (e.g., Terra’s UST/LUNA, AMPL).
    \end{itemize}
    \item No Yield: Assets that do not generate yield by themselves, functioning only as stores of value or means of payment. (e.g., BTC, LTC) 
\end{itemize}

\paragraph{Distribution Mechanism}
Yield-bearing tokens can be further distinguished by how yield is operationalised within the protocol. The mechanism determines the way that yield is reflected in tokenholder value. The distribution mechanism defines the structure of the token holder’s claim and the degree of control retained by the issuer or protocol, which dictates the necessary \ac{tradfi} analogy and the level of required prudential oversight.

\begin{itemize}[noitemsep]
    \item Quantity-accrual (rebasing): Yield is distributed by periodically adjusting the holder’s balance upward (e.g., stETH, AMPL). The mechanism of automatically increasing the token balance is analogous to a Pass-Through Certificate in traditional finance. From a legal-structural perspective, this implies the holder receives an undivided, pro-rata share of the cash flows from the underlying pool of assets. This structure generally supports a more decentralised, hands-off legal profile, consistent with the low-risk \enquote{administrative} function under US law.
    \item Value-accrual: Token quantity remains constant, but its redemption or reference ratio to the underlying asset increases over time (e.g., rETH, cbETH, BETH). When a token uses the Value-Accrual mechanism, where the quantity held is constant, but the token's reference value increases over time as yield is capitalised, it functionally resembles a Capitalising Share Class in traditional funds. This design often requires the issuer to retain centralised control over the accounting, fee application, and the increasing redemption rate. This concentration of control results in a centralised classification, which, in turn, could potentially satisfy the managerial effort requirement of the US Howey test, thereby elevating the asset's security risk.
    \item Price-accrual: Neither token balance nor redemption rate changes; instead, yield is distributed through price adjustment (e.g., mSOL).
    \item External redistribution: Yield generated by a protocol is distributed in another asset or via a distinct rewards pool (e.g., AnkrETH).
\end{itemize}

In essence, while the Yield Source helps determine if the asset is a security or not, the Distribution Mechanism defines the type of financial instrument it is, dictating the specific legal obligations (e.g., custody, capitalisation, liability) imposed on the issuer.

\subsubsection{Redemption Type}
The economic purpose of tokens can be distinguished by their link to reference assets. This aligns with regulatory categories under the \ac{MiCAR} and the U.S. GENIUS Act.  

Within this framework, redemption reference asset can be organised as follows:

\paragraph{Reference Asset}
\ac{MiCAR} distinguishes between \ac{emt}s and \ac{art}s:  \footnote{\ac{MiCAR} defines e-money tokens \enquote{A type of crypto-asset that purports to maintain a stable value by referencing the value of one official currency.} and \ac{art}s \enquote{a type of crypto-asset that is not an electronic money token and that purports to maintain a stable value by referencing another value or right or a combination thereof, including one or more official currencies}. (See above definitions in Regulation (EU) 2023/1114, \url{www.eba.europa.eu/regulation-and-policy/asset-referenced-and-e-money-tokens-mica}) }. 
The U.S. GENIUS Act defines a unified category of \enquote{payment stablecoins,} exempting them from securities law if they meet reserve and redemption criteria.\footnote{CRS Insight IN12522: The Act defines a \enquote{payment stablecoin} as a digital asset designed to maintain a stable value relative to a reference asset and redeemable for monetary value at par. The statute provides that such instruments shall not be treated as securities if compliant with reserve and redemption requirements.}  

Within this framework, redemption reference asset can be organised as follows:

\begin{itemize}[noitemsep]
    \item Fiat-pegged stablecoins: \ac{MiCAR} \ac{emt} in the EU and payment stablecoins under the GENIUS Act (U.S.).
    \begin{itemize}[noitemsep]
        \item Par-for-reserve: (e.g., USDC, USDT).  
    \end{itemize}

    \item \ac{art}s: assets that purport to maintain a stable value by referencing one or more assets, and designed to defend value parity through issuer redemption or collateral mechanisms.
    \begin{itemize}[noitemsep]
        \item Physical redemption: (e.g., PAXG).   
        \item Collateral redemption: Redeem against posted collateral (e.g., LUSD, DAI) as implemented in protocols.  
    \end{itemize}

    \item Non-stablecoins with redemption features: crypto-referenced tokens not intended to stabilise their own value (hence distinct from \ac{art}s). 
    \begin{itemize}[noitemsep]
        \item Wrapped tokens: Burn-and-release (e.g., wBTC, cbETH). 
        \item \ac{lst}s: Validator withdrawal (e.g., rETH, stATOM, mSOL).
    \end{itemize}
    \item Non-stablecoins without redemption: Assets providing no redemption right. (e.g., BTC, ETH, UNI, etc)

\end{itemize}

\paragraph{Redemption mechanism}
Assets can be further distinguished by how redemption is operationalised within the protocol. The mechanism determines how a holder converts the token into its reference asset, shaping liquidity, run dynamics, and settlement risk. The mechanism of redemption is crucial because it reveals where control resides and how market stress is managed, directly impacting US security classification and EU prudential concerns.

\begin{itemize}[noitemsep]
    \item Off-chain issuer redemption: Issuer or custodian redeems off chain (e.g., USDC, USDT). Requires the issuer or custodian to unilaterally authorize and execute the release of the reserve asset (fiat). This centralised control satisfies the \enquote{managerial efforts} prong of the Howey test , meaning the asset is legally dependent on the issuer’s continuous managerial function. This central control creates counterparty risk and exposes the asset to potential seizure or control by a single entity, which is a key regulatory concern for large, centrally-managed stablecoins.
    \item  Burn-to-unlock: Holder repays or burns and withdraws collateral from protocol custody (e.g., DAI (maker vaults)). This is programmatic control, where the right to redemption is enforced by a smart contract. This programmatic control may be critical under the US Howey Test, as it may be instrumental to demonstrate the absence of centralised, human \enquote{managerial effort} required to free the collateral, thereby helping to mitigate the risk of security classification. Furthermore, the underlying economic function, the locking of collateral to mint a token (debt) and the subsequent burning of the token to retrieve the collateral allows regulators to map this crypto activity to established traditional finance instruments, such as a Margin Loan or a Repurchase Agreement (Repo), which aids in applying familiar prudential frameworks for risk assessment.
    \item Bridge burn-and-release: Burn or lock a wrapped asset to release the underlying on another chain (e.g., WBTC). Requires reliance on a centralised custodian (e.g., BitGo) to lock the underlying asset and execute the burn/release.
    \item Secondary-market exit: holder exits by swapping to others (e.g., cbETH, BETH). Reinforces a Centralised classification due to issuer discretion over capitalised returns.
    \item Queued withdrawal: Redemption subject to queues or withdrawal windows preestablished in the protocol (e.g., stETH, rETH).
    \item Protocol par redemption: Built-in facility to swap at or near par using protocol reserves or troves (e.g., DAI (maker \ac{psm})).
    \item  Claim-from-pool: Burn LP or position token to withdraw a pro-rata share of pool reserves (e.g., cDAI).
\end{itemize}

The type of redemption mechanism also has prudential implications: off-chain issuer redemption (custodial) may trigger licensing and safeguarding obligations under MiCAR Titles III and IV, while on-chain mechanisms (e.g., burn-to-unlock, bridge burn-and-release) introduce operational, liquidity, and settlement risks that regulators could associate with market-infrastructure oversight (see ESMA Consultation on MiCAR Technical Standards, 2024).

Under U.S. law, redemption is equally decisive. As mentioned in the GENIUS Act, distinguish between payment stablecoins and investment-type tokens hence if a token offers redemption backed by reserves and no expectation of profit, it is typically treated as a payment instrument rather than an investment contract.
In short, redemption type determines whether a crypto-asset creates a legal claim, a regulatory obligation, or merely a market exposure. It is therefore essential for establishing the boundary between payment, investment, and commodity classifications across jurisdictions.

\subsection{Legal Dimension}
This dimension situates each crypto-asset within existing regulatory and private-law concepts. Firstly, the legal classification subdimension show how assets are categorised and thereby treated under current EU and US regulation framework. Then claim form subdimension distinguishes assets according to economic rights embedded in assets along three orthogonal axes: claim form, claim target, direction of proceeds. 

\subsubsection{Legal Classification}

This dimension captures how a crypto-asset is positioned within the current US and EU legal frameworks. It is a descriptive label derived from the US Howey investment contract test, the EU \ac{mifid} financial instrument and \ac{MiCAR}/AIFMD (ART-AIF) tests.
The goal is to indicate whether an asset is currently treated as a security, financial instrument, a stable-value instrument (\ac{art}/\ac{emt}), an AIF-type fund share, or as \enquote{other} crypto-asset.

Operationally, we think of these tests as decision procedures over the following elements:

\begin{itemize}
    \item US Howey test: Checks whether there is (i) investment of money, (ii) a common enterprise, (iii) an expectation of profits, and (iv) profits that depend on entrepreneurial or managerial efforts that are not the investor's own ( efforts from the promoter's, from an identifiable 3rd party or not an identifiable 3rd party).
    \item EU \ac{mifid} financial-instrument test\footnote{EU MiFID financial-instrument test for crypto-assets consists in checking whether the token falls into any of the financial-instrument classes in Annex I Section C MiFID II, most commonly transferable securities (point (1)), units or shares in a collective investment undertaking (point (3)), or derivative contracts (points (4)–(10))}: Checks whether the asset grants (i) profits or repayment of capital, and (ii) a legal claim against an identifiable issuer.
    \item EU ART-AIF\footnote{Alternative Investment Fund (AIF) defined in AIFMD (Directive 2011/61/EU) Article 4(1)(a)} test: Checks whether there is (i) pooled risk and return, (ii) a defined investment policy, (iii) for the benefit of those investors.
\end{itemize}

With these tests, we categorize assets to 4 main categories:
\begin{itemize}
    \item Security or financial instrument: Assets that at least meet one of the US Howey test and EU \ac{mifid} test (e.g., EXIT, INX).
    \item Stable-value token (\ac{art}/\ac{emt}): Assets fail US Howey test and EU \ac{mifid} test, but pass \ac{MiCAR} stablecoin test, whose primary purpose is to maintain a stable value relative to one or more reference assets. Typical examples are fiat-backed stablecoins (e.g., USDC, USDT), overcollaterised or basket-backed stablecoins (e.g., DAI).
    \item Fund / AIF-type: Assets satisfy Howey test, and trigger EU AIF classification, assets that represent units or shares in a collective investment undertaking or alternative investment fund in the EU sense. Include tokenised fund shares (e.g., SPICE, BUIDL, ACRED).
    \item Other crypto-assets or commodities: Assets do not qualify any classification mention above and fail all three test. Legal claims are weak or absent; holders typically have only in rem claims on their own tokens, not on a reserve or issuer balance sheet. Include native tokens (e.g., BTC) or pure governance/utility tokens (e.g., MKR, COMP).
\end{itemize}

This dimension is intended to be descriptive and orthogonal to the economic-rights and redemption dimensions. Assets categorised as security or financial instrument fall within securities law in US and thus their registration or exemptions under the Securities Act, ongoing disclosure and market-abuse rules, and trading via registered intermediaries; in EU they fall outside of \ac{MiCAR} and instead governed by the \ac{mifid}/Prospectus/AIFMD/UCITS framework. Assets categorised as stable-value tokens are typically treated as commodities or payment instruments in US, and are in-scope of \ac{MiCAR} \ac{art}s or \ac{emt}s, this categorisation indicates that regulatory attention focuses on reserve adequacy, redemption rights, and systemic risk, rather than on investor-protection rules for investment products. Fund / AIF-type assets are both a security and a collective investment vehicle unit, making manager-level regulation central to the risk assessment.

\subsubsection{Form of Claim}

Crypto-assets can embed very different positions for their holders. Some merely provide a contractual claim against an issuer or platform, while others purport to give a property-type interest in a segregated pool of assets or in the on-chain object itself. In this dimension, we distinguish the form of rights along three orthogonal axes: (i) whether the holder has an in personam (contractual) or in rem (property-type) position, (ii) whether the claim is directed primarily against the issuer’s balance sheet or against a specified reserve or asset pool, (iii) whether the proceeds are retailed by holders or by issuers (or no proceeds). The proceeds to issuer and no proceeds are collapsed into one because from the token-holder’s perspective, both cases imply the same thing: holders have no enforceable claim to any cash flows generated by the underlying assets. Since an in rem position, by construction, presupposes a right in a defined pool of assets, we restrict in rem to claims against a reserve, which yields six economically relevant combinations.

\begin{itemize}
    \item In personam, against issuer, proceeds to issuer: The holder has a contractual or statutory claim against the issuer or platform, but no direct right to specific backing assets, and the issuer keeps the interest margin (e.g., USDC, USDT).
    \item In personam, against issuer, proceeds to holders: Holders have equity or profit-sharing claims on the issuing entity, such as tokenised euity (e.g., TZROP).
    \item In personam, against reserve, proceeds to issuer: The holder’s right is still formulated as a claim for performance, but explicitly linked to a segregated pool of backing asset reserve; holders are not entitled to interest on the reserve,  the issuer captures the interest margin as revenue (e.g., PYUSD).
    \item In personam, against reserve,  proceeds to holders: contractual claims on that pool’s cash flows are allocated to tranche holders.
    \item In rem, against reserve, proceeds to issuer: The asset represents co-ownership or a beneficial interest in a defined asset pool, net income accrues to investors. (e.g., PAXG)
    \item No claim: Assets confer no economic claim. (e.g., DOGE, BTC)
\end{itemize}

The form of rights does not by itself determine whether a crypto-asset qualifies as a \ac{mifid} financial instrument or a \ac{MiCAR} token. However, distinguishing in personam from in rem positions, and claims against an issuer from claims against a segregated reserve, is essential for assessing insolvency risk, investor protection, and the degree to which holders are exposed to issuer credit risk rather than asset-pool performance.

 Building on the taxonomy dimensions outlined above, we represent each asset as a Python @dataclass CryptoAsset, whose fields are typed with small Literals/Booleans drawn from our dimensional categories. We encode classification-relevant properties as controlled facets. Given any new asset, its feature tuple deterministically maps to a category, enabling reproducible classification and consistent regulation.

%% file: sections/dataclass.tex
\section{Taxonomy Dataclass and Decision Rules}

In this part, we show how crypto assets are mapped in our taxonomy structure. We collect features of top 100 assets (by market cap) snapshot from CoinGecko at 19 Nov 2025, derive their categorisation on centralisation dimension, and where applicable, map assets into \ac{tradfi} analogies with decision rules defined in \autoref{fig:analogies}. The categorising features are manually collected from project documents, protocol whitepapers, and legal text. For number of parties in control under market category as defined in \autoref{sec: centralisation_dimension}, we fetch top holder and balance from Bitquery. Constrained by the free key call limit, we consider on-chain holders only on Ethereum and top 100 holders, when one holder controls over 60\% among top 100 balance, ownership makeup is considered to be controlled by only one party. 

\subsection{Dataclass Design}
\label{sec:dataclass_design}
We encode the taxonomy in a single dataclass (CryptoAsset) that inherits Asset dataclass, Asset has attributes of issuer and \enquote{is\_fiat}. Instantiating CryptoAsset, explicit features are provided at initiation, while derived features are computed deterministically by property methods that implement our decision rules. Initiate inputs are typed as strings, Literals, booleans and dictionary.

\begin{figure}

    \centering
    \includegraphics[width=0.5\linewidth]{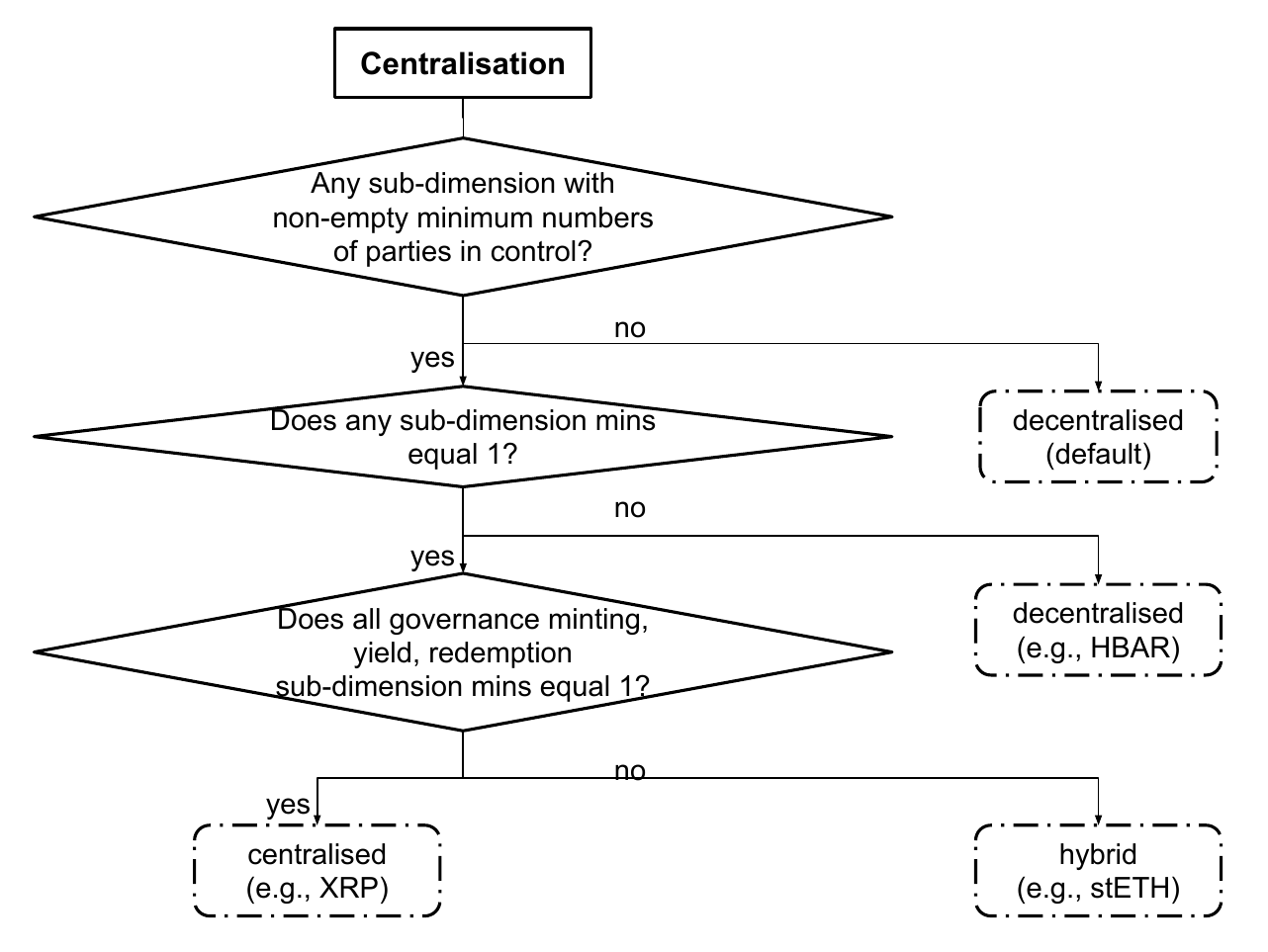}
    \caption{Centralisation Decision Tree}
    \label{fig:centralisation}
\end{figure}
\begin{figure}
    \centering
    \includegraphics[width=0.5\linewidth]{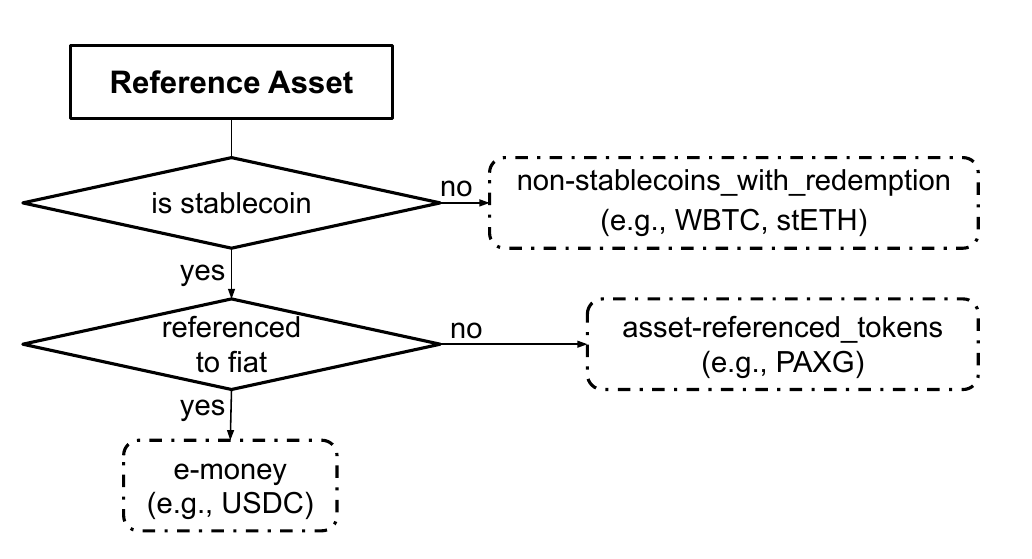}
    \caption{Reference Asset Decision Tree}
    \label{fig:reference_asset}
\end{figure}

To derive an crypto asset’s centralisation label, we encode the centralisation surface of each asset as one nested dictionary for eight sub-dimensions listed in \autoref{tab: critical_resources}, this surface maps each sub-dimension to a dictionary with form \{resource\_i: j\}, where resource\_i denotes critical resources in this sub-dimension and $j \in N$ is the minimum number of independent parties required to exercise effective control. Keys may be omitted and all values default to None when not applicable or no bound number of entities can be identified. The asset’s position on the centralisation dimension is then computed by evaluating each sub-dimension for unilateral control (j=1) and aggregating the eight outcomes into the asset-level classification, see decision rules in \autoref{fig:centralisation}. 

To derive reference\_asset categories, we input reference asset (Asset instance) and \enquote{is\_stablecoin} (boolean), the categorisation follows rules in \autoref{fig:reference_asset}.

\begin{figure}
    \centering
    \includegraphics[width=0.9\linewidth]{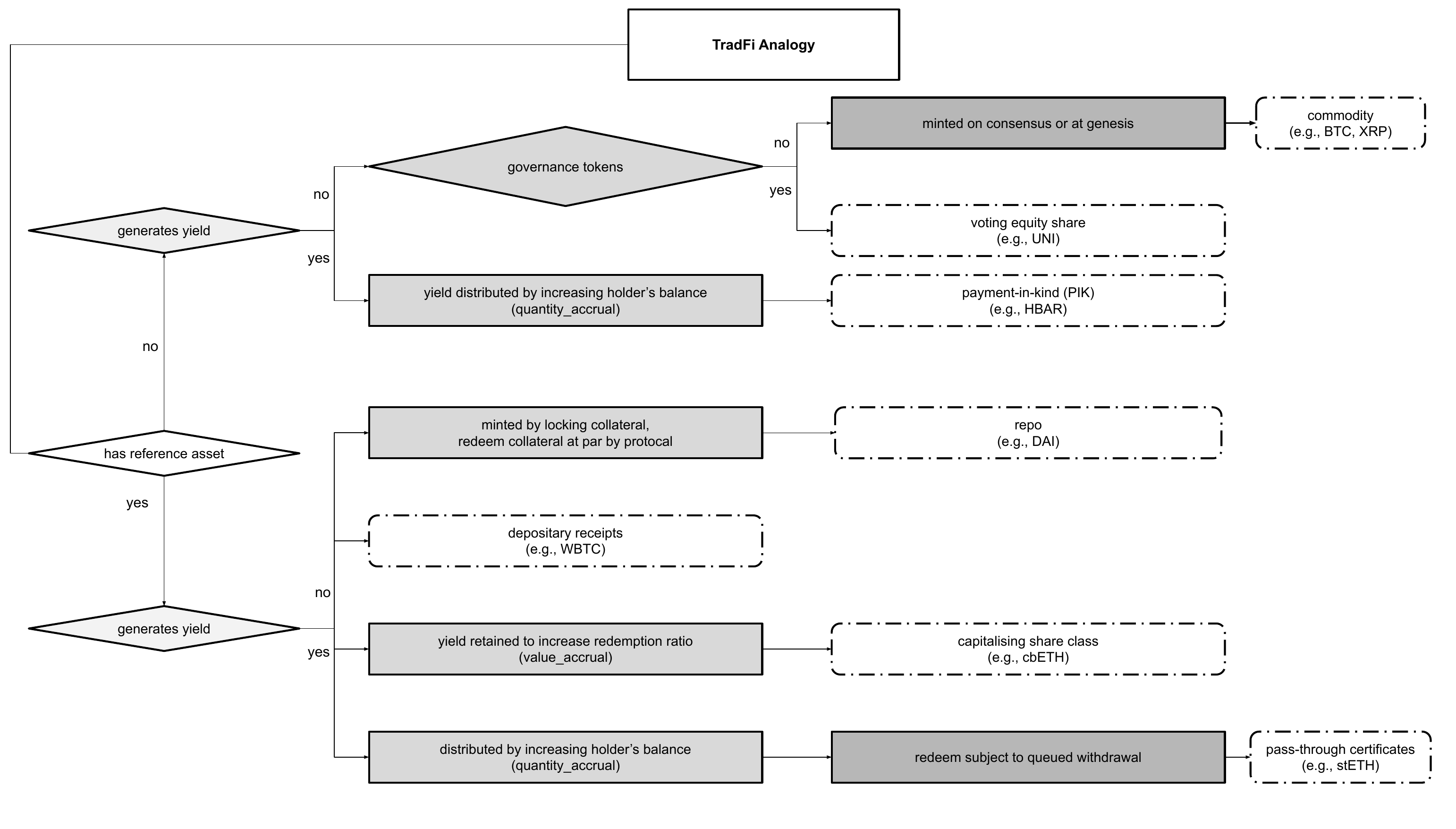}
    \caption{TradFi analogy Decision Tree}
    \label{fig:analogies}
\end{figure}

To make the taxonomy more interpretable for regulators and investors, we derive \ac{tradfi} analogies (to commodities, voting equity shares, payment in kind, repo, depositary receipts, capitalising share class, and \ac{ptc}s) for matching combinations of features. Linking the risk profile and design to familiar financial instrument buckets, we show where existing regulation might be relevant. For example, when one asset generates yield that's accrued by increasing holders' balance, and doesn't have any reference asset, its economic and functional features are analogous to payment in kind, this similarity allows us to refer to well defined \ac{tradfi} instruments when regulating or investing this asset. These analogies are exposed as read-only properties, the mapping is deterministic and auditable. With all the explicit features and derived features, we print one analogy for each crypto asset, these analogies are mutually exclusive, when no defined analogy found, it's set to default \enquote{other}. See the analogy decision tree in \autoref{fig:analogies}.

\subsection{Mapping Results}
\begin{figure}[htbp]
    \centering
    \subfloat[Structural and constitutive design\label{fig:parallel_fund}]{\includegraphics[width=\linewidth]{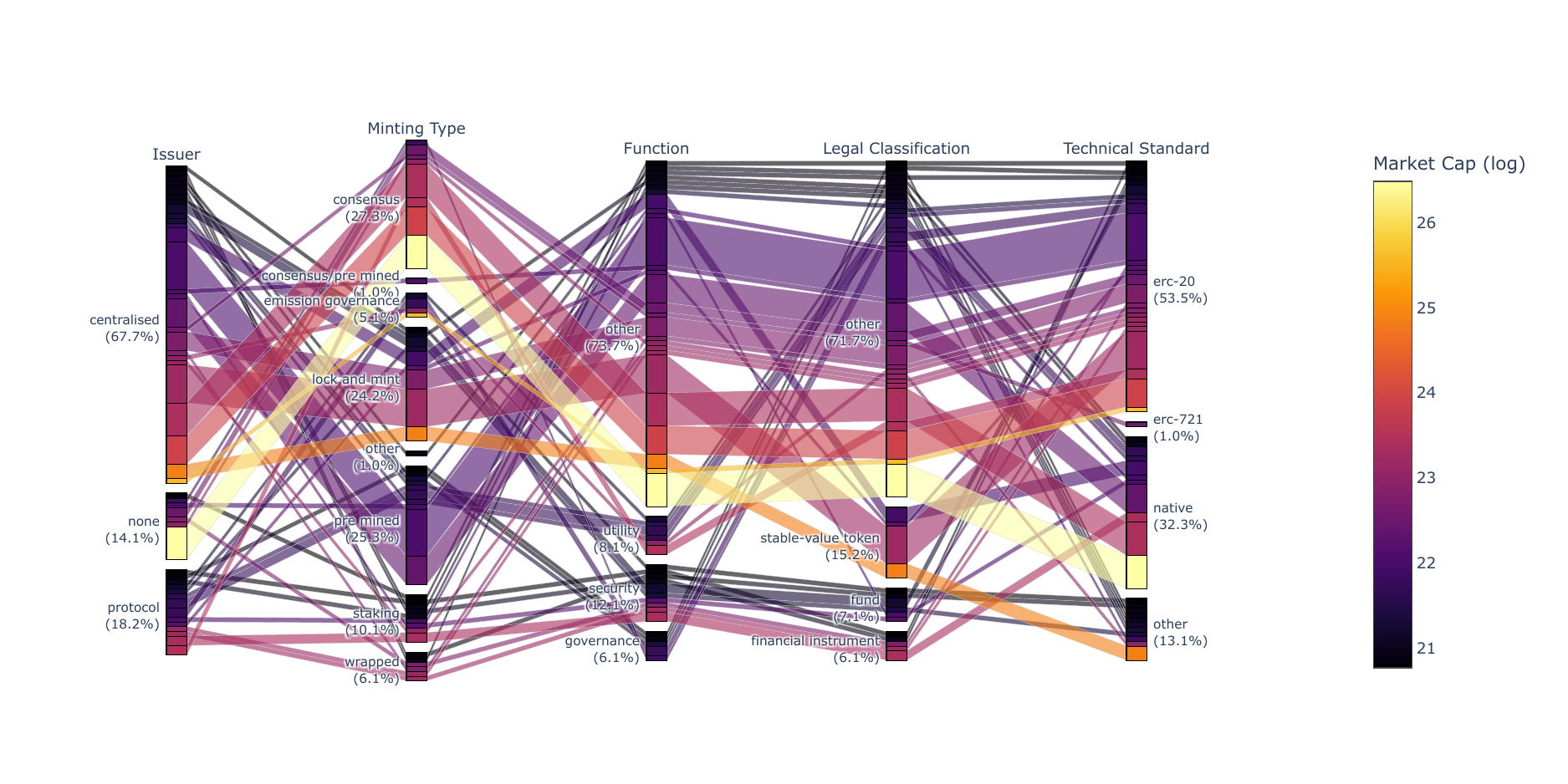}}
    \vspace{0.2em}
    \subfloat[Economic and value-realisation\label{fig:parralel_eco}]{\includegraphics[width=\linewidth]{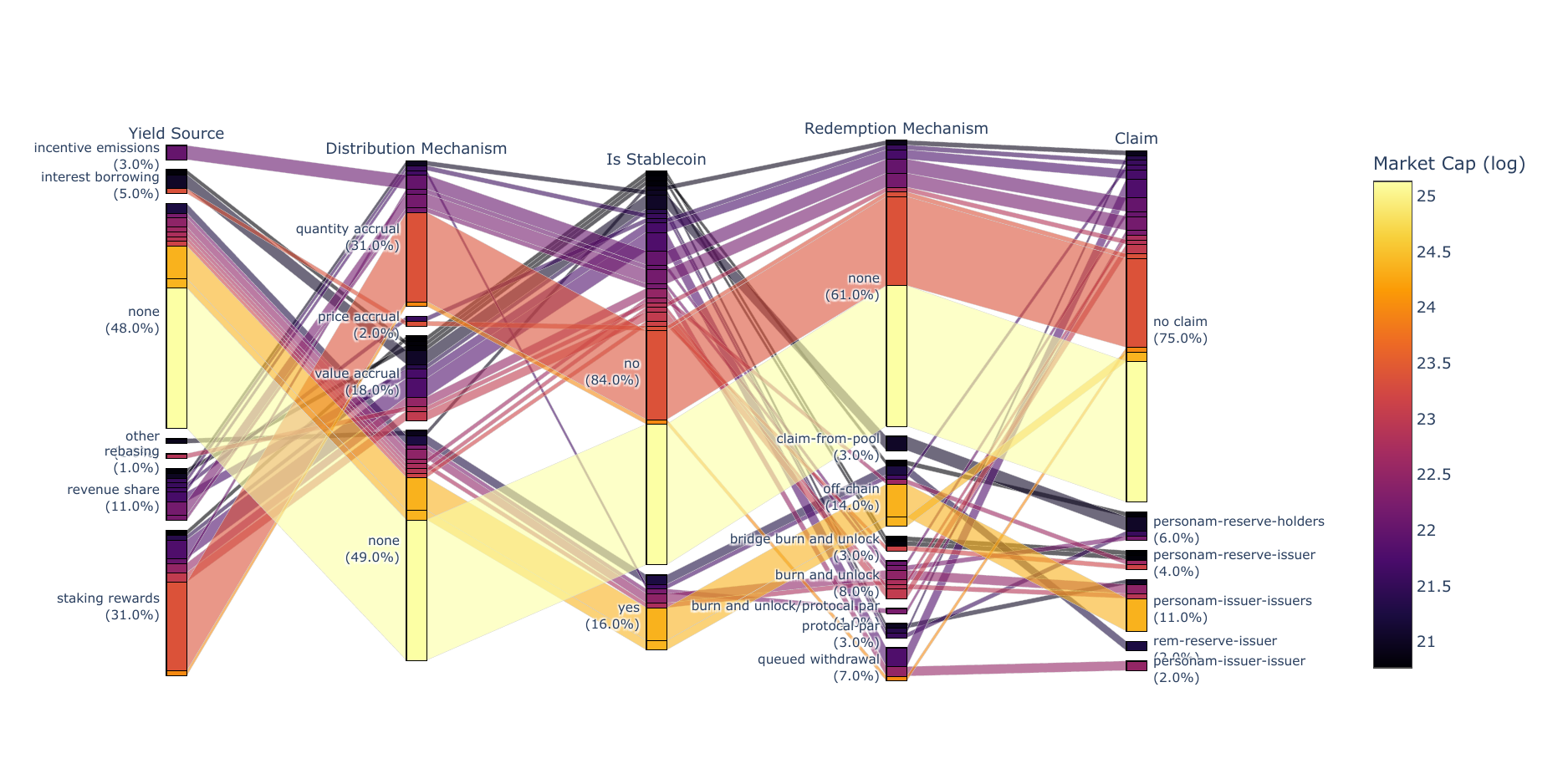}}

    \caption{Distribution of Explicit Features for Top 100 Assets. The top figure shows distribution of structural and constitutive design features: issuer, minting type, function, legal classification, and technical standard. The bottom figure shows distribution of economic and value-realisation features: yield source, distribution mechanism, stablecoin, redemption mechanism and form of claim. For compactness, we use notation {personam/rem–reserve/issuer–issuer/holder} for form of legal claim, where each position corresponds to claim in personam/rem, against reserve/issuer, proceeds to issuer/holder. 
    The width of each path represents the number of assets in that path, while the colour indicates the logarithm of the mean market capitalisation of those assets as of 19 November 2025.}
    \label{fig:explicit_features}
\end{figure}

We manually collected all explicit features for top 100 assets, and instantiate each asset with CryptoAsset mentioned in \autoref{sec:dataclass_design}, and print \ac{tradfi} analogies. See the distribution of explicit features in \autoref{fig:explicit_features}. 
Over half of the top assets are implemented on ERC-20 standard (54\%), and 32\% of the assets are implemented on native standard. The prevalence of ERC-20 enhances composability by standardising how assets are transferred and integrated into \ac{defi} protocols. From a regulatory perspective, this standardisation may support more unified, technically grounded requirements. Most assets are classified as having a \enquote{other} function, including certain assets (e.g., stablecoins) whose functional role is not always explicitly stated or consistently recognised. This concentration highlights the need for a more granular and economically grounded approach to function identification, particularly for crypto-assets whose design and use cases do not map cleanly onto \ac{tradfi} instruments or conventional product categories.

Among the 100 assets, 68 are issued by an identifiable (centralised) issuer. The dominant minting mechanisms (lock-and-mint, pre-mined issuance, and consensus-based minting) are, in practice, more frequently associated with issuer-led control over issuance and supply management. While many crypto-assets are commonly described as “decentralised”, this distribution indicates that asset issuance is often materially centralised, which has direct regulatory relevance. In particular, concentrated issuance authority can affect accountability, disclosure expectations, governance oversight, and the feasibility of enforcing conduct and prudential requirements.

50 assets bear yield, among which 62\% are accrued through quantity accrual and 34\% are value accrual. Moreover, for the majority of yield-bearing assets (82\%), returns arise from revenue-sharing arrangements or staking rewards. This concentration suggests that yield generation in crypto-asset markets is predominantly tied to protocol-level cash-flow distribution or participation incentives.

In total, 37 assets are referenced to other assets (and 16 among them are stablecoins). Redemption mechanisms are relatively evenly represented, with no single mechanism materially dominating the sample. MiCAR provides a relatively comprehensive framework for stablecoins, but the market is increasingly innovating in reference-based non-stablecoin designs, effectively moving activity toward instruments that may sit outside the stablecoin rules. This development shifts the regulatory and investment focus from “stablecoins” per se to the broader set of assets that embed reference mechanisms and associated risks.

According to the definition in \autoref{sec: centralisation_dimension}, 45 assets from the top 100 are categorised as decentralised, 39 assets categorised as centralised, and 8 as hybrid, see mapped centralisation buckets in \autoref{tab: centralisation}. Some of the assets are usually categorised and marketed as decentralised, such as XRP, LBTC, WLD. Though they are issued and transacted via smart contracts, critical powers such as upgrading contracts, changing parameters, pausing markets are controlled by single entity, therefore in our taxonomy classify such assets as centralised. 
Following the decision rules in \autoref{fig:analogies}, top 100 assets are mapped to \ac{tradfi} analogies, the mapped analogy buckets are shown in \autoref{tab: analogue}.


\input{tables/tradfi_table}
\begin{flushleft}
\footnotesize
\textit{Note:}  
The table reports the top 100 crypto-assets by market capitalisation as of 19 Nov 2025.  
One additional asset (\textit{HBARX}) is included as a case-relevant example.
\end{flushleft}
\input{tables/centralisation_table}
\begin{flushleft}
\footnotesize
\textit{Note:}  
The table reports the top 100 crypto-assets by market capitalisation as of 19 Nov 2025.  
One additional asset (\textit{HBARX}) is included as a case-relevant example.
\end{flushleft}

%% file: tables/tradfi_table.tex
\begin{longtable}{p{0.20\textwidth} p{0.80\textwidth}}
\caption{TradFi analogues for the top 100 assets}\label{tab: analogue}\\
\toprule
Tradfi\_analogy & Assets \\
\midrule
\endfirsthead

\toprule
Tradfi\_analogy & Assets \\
\midrule
\endhead

\bottomrule
\endfoot

\bottomrule
\endlastfoot

capitalising share class & \begin{tabular}{@{}ccccccc@{}}
\begin{tabular}{@{}c@{}}\includegraphics[height=4ex]{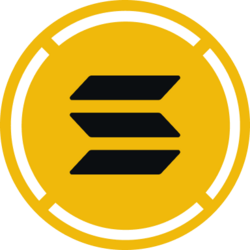}\\\footnotesize BNSOL\end{tabular} & \begin{tabular}{@{}c@{}}\includegraphics[height=4ex]{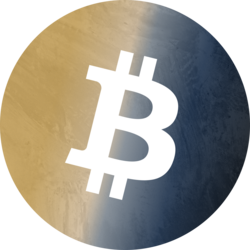}\\FBTC\end{tabular} & \begin{tabular}{@{}c@{}}\includegraphics[height=4ex]{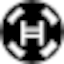}\\\footnotesize HBARX\end{tabular} & \begin{tabular}{@{}c@{}}\includegraphics[height=4ex]{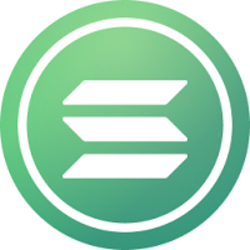}\\\footnotesize JitoSOL\end{tabular} & \begin{tabular}{@{}c@{}}\includegraphics[height=4ex]{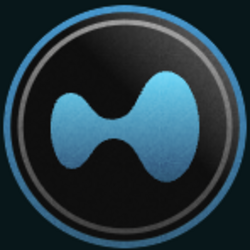}\\\footnotesize KHYPE\end{tabular} & \begin{tabular}{@{}c@{}}\includegraphics[height=4ex]{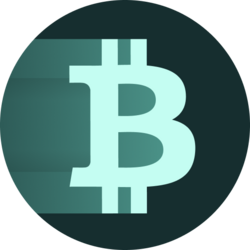}\\LBTC\end{tabular} & \begin{tabular}{@{}c@{}}\includegraphics[height=4ex]{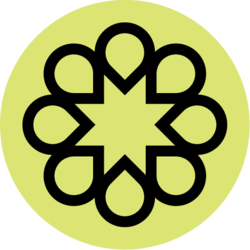}\\\footnotesize LsETH\end{tabular} \\
\begin{tabular}{@{}c@{}}\includegraphics[height=4ex]{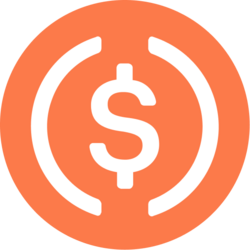}\\\footnotesize SYRUPUSDC\end{tabular} & \begin{tabular}{@{}c@{}}\includegraphics[height=4ex]{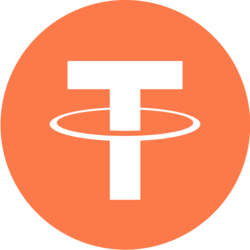}\\\footnotesize SYRUPUSDT\end{tabular} & \begin{tabular}{@{}c@{}}\includegraphics[height=4ex]{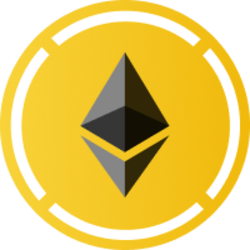}\\\footnotesize WBETH\end{tabular} & \begin{tabular}{@{}c@{}}\includegraphics[height=4ex]{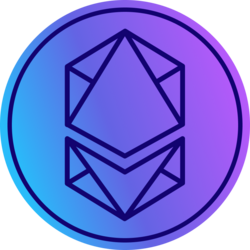}\\\footnotesize WEETH\end{tabular} & \begin{tabular}{@{}c@{}}\includegraphics[height=4ex]{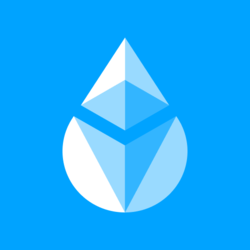}\\\footnotesize WSTETH\end{tabular} & \begin{tabular}{@{}c@{}}\includegraphics[height=4ex]{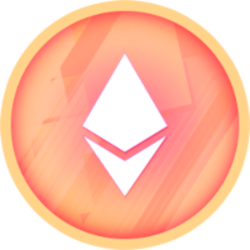}\\rETH\end{tabular} &  \\
\end{tabular} \\[0.5em]
commodity & \begin{tabular}{@{}ccccccc@{}}
\begin{tabular}{@{}c@{}}\includegraphics[height=4ex]{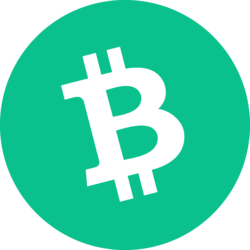}\\BCH\end{tabular} & \begin{tabular}{@{}c@{}}\includegraphics[height=4ex]{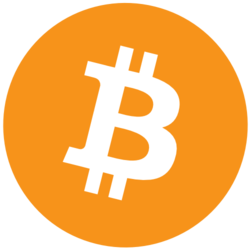}\\BTC\end{tabular} & \begin{tabular}{@{}c@{}}\includegraphics[height=4ex]{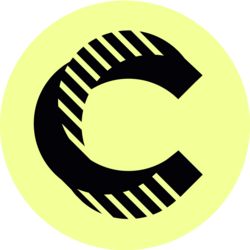}\\CC\end{tabular} & \begin{tabular}{@{}c@{}}\includegraphics[height=4ex]{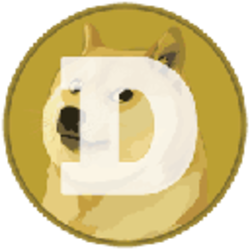}\\DOGE\end{tabular} & \begin{tabular}{@{}c@{}}\includegraphics[height=4ex]{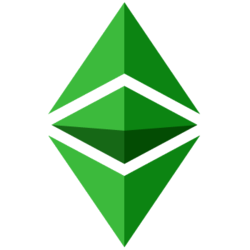}\\ETC\end{tabular} & \begin{tabular}{@{}c@{}}\includegraphics[height=4ex]{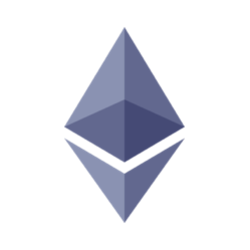}\\ETH\end{tabular} & \begin{tabular}{@{}c@{}}\includegraphics[height=4ex]{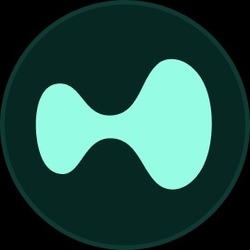}\\HYPE\end{tabular} \\
\begin{tabular}{@{}c@{}}\includegraphics[height=4ex]{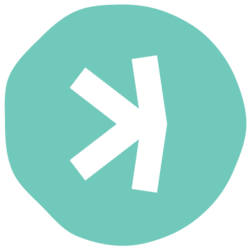}\\KAS\end{tabular} & \begin{tabular}{@{}c@{}}\includegraphics[height=4ex]{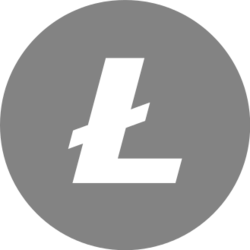}\\LTC\end{tabular} & \begin{tabular}{@{}c@{}}\includegraphics[height=4ex]{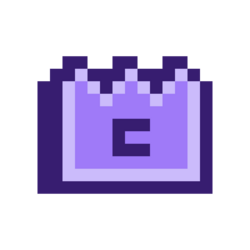}\\M\end{tabular} & \begin{tabular}{@{}c@{}}\includegraphics[height=4ex]{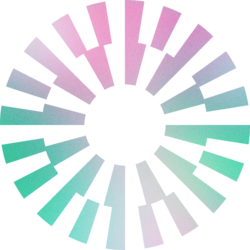}\\MNT\end{tabular} & \begin{tabular}{@{}c@{}}\includegraphics[height=4ex]{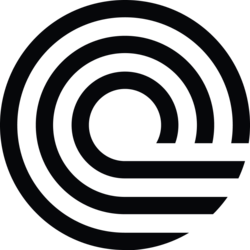}\\ONDO\end{tabular} & \begin{tabular}{@{}c@{}}\includegraphics[height=4ex]{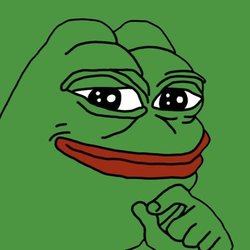}\\PEPE\end{tabular} & \begin{tabular}{@{}c@{}}\includegraphics[height=4ex]{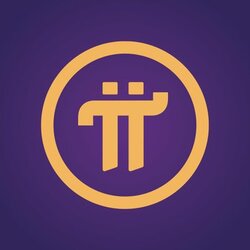}\\PI\end{tabular} \\
\begin{tabular}{@{}c@{}}\includegraphics[height=4ex]{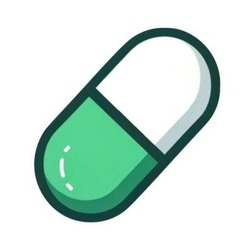}\\PUMP\end{tabular} & \begin{tabular}{@{}c@{}}\includegraphics[height=4ex]{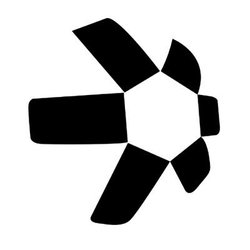}\\QNT\end{tabular} & \begin{tabular}{@{}c@{}}\includegraphics[height=4ex]{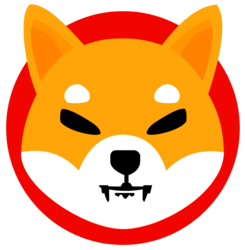}\\SHIB\end{tabular} & \begin{tabular}{@{}c@{}}\includegraphics[height=4ex]{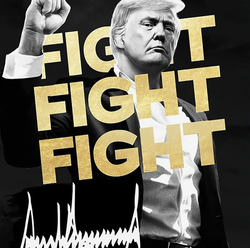}\\\footnotesize TRUMP\end{tabular} & \begin{tabular}{@{}c@{}}\includegraphics[height=4ex]{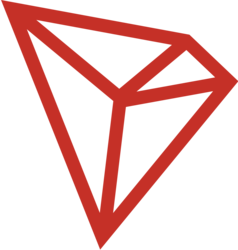}\\TRX\end{tabular} & \begin{tabular}{@{}c@{}}\includegraphics[height=4ex]{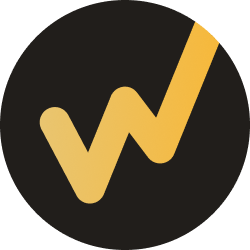}\\WBT\end{tabular} & \begin{tabular}{@{}c@{}}\includegraphics[height=4ex]{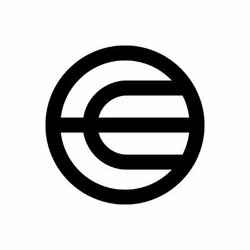}\\WLD\end{tabular} \\
\begin{tabular}{@{}c@{}}\includegraphics[height=4ex]{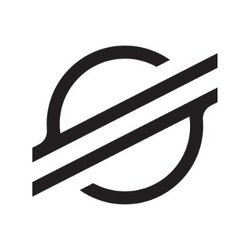}\\XLM\end{tabular} & \begin{tabular}{@{}c@{}}\includegraphics[height=4ex]{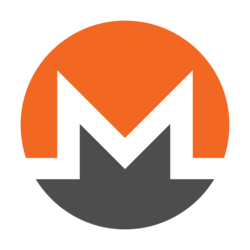}\\XMR\end{tabular} & \begin{tabular}{@{}c@{}}\includegraphics[height=4ex]{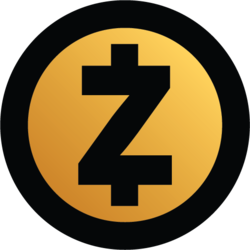}\\ZEC\end{tabular} &  &  &  &  \\
\end{tabular} \\[0.5em]
depositary receipt & \begin{tabular}{@{}ccccccc@{}}
\begin{tabular}{@{}c@{}}\includegraphics[height=4ex]{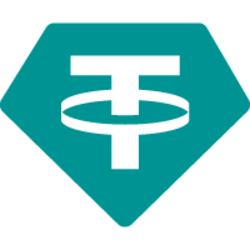}\\\footnotesize BSC-USD\end{tabular} & \begin{tabular}{@{}c@{}}\includegraphics[height=4ex]{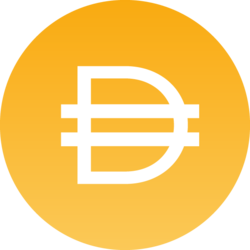}\\DAI\end{tabular} & \begin{tabular}{@{}c@{}}\includegraphics[height=4ex]{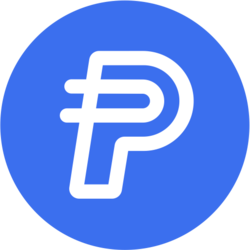}\\\footnotesize PYUSD\end{tabular} & \begin{tabular}{@{}c@{}}\includegraphics[height=4ex]{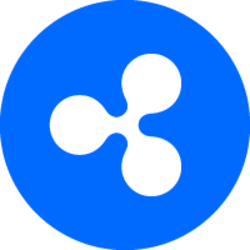}\\\footnotesize RLUSD\end{tabular} & \begin{tabular}{@{}c@{}}\includegraphics[height=4ex]{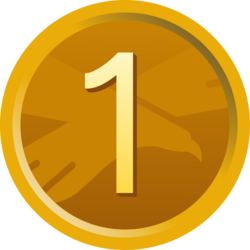}\\USD1\end{tabular} & \begin{tabular}{@{}c@{}}\includegraphics[height=4ex]{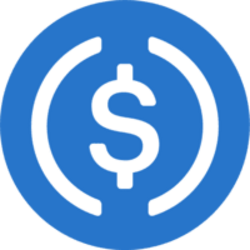}\\USDC\end{tabular} & \begin{tabular}{@{}c@{}}\includegraphics[height=4ex]{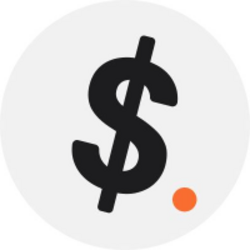}\\USDF\end{tabular} \\
\begin{tabular}{@{}c@{}}\includegraphics[height=4ex]{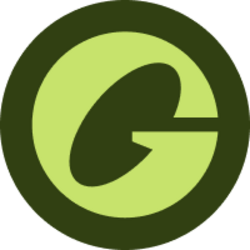}\\USDG\end{tabular} & \begin{tabular}{@{}c@{}}\includegraphics[height=4ex]{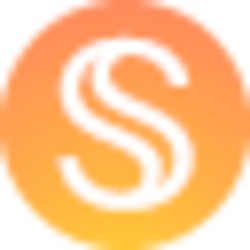}\\USDS\end{tabular} & \begin{tabular}{@{}c@{}}\includegraphics[height=4ex]{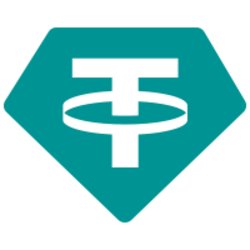}\\USDT\end{tabular} & \begin{tabular}{@{}c@{}}\includegraphics[height=4ex]{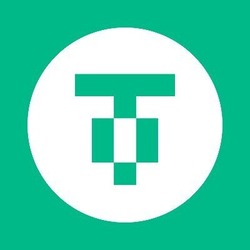}\\\footnotesize USDT0\end{tabular} & \begin{tabular}{@{}c@{}}\includegraphics[height=4ex]{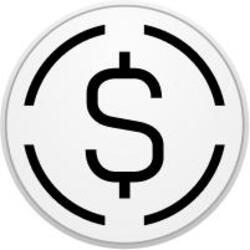}\\\footnotesize USDTB\end{tabular} & \begin{tabular}{@{}c@{}}\includegraphics[height=4ex]{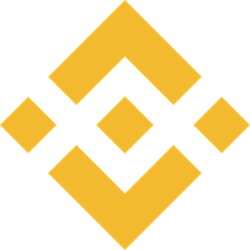}\\WBNB\end{tabular} & \begin{tabular}{@{}c@{}}\includegraphics[height=4ex]{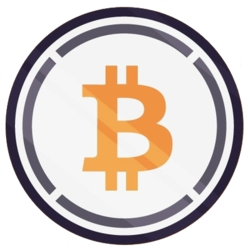}\\WBTC\end{tabular} \\
\begin{tabular}{@{}c@{}}\includegraphics[height=4ex]{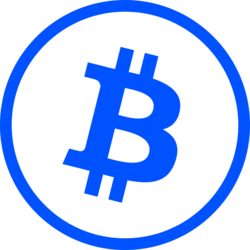}\\\footnotesize cbBTC\end{tabular} &  &  &  &  &  &  \\
\end{tabular} \\[0.5em]
other & \begin{tabular}{@{}ccccccc@{}}
\begin{tabular}{@{}c@{}}\includegraphics[height=4ex]{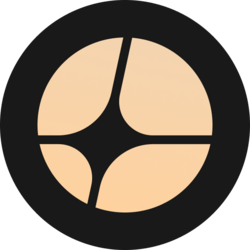}\\\footnotesize ASTER\end{tabular} & \begin{tabular}{@{}c@{}}\includegraphics[height=4ex]{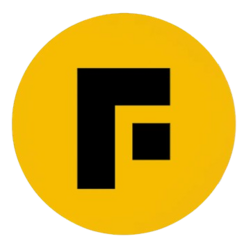}\\\footnotesize BFUSD\end{tabular} & \begin{tabular}{@{}c@{}}\includegraphics[height=4ex]{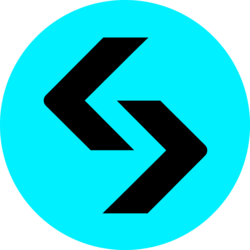}\\BGB\end{tabular} & \begin{tabular}{@{}c@{}}\includegraphics[height=4ex]{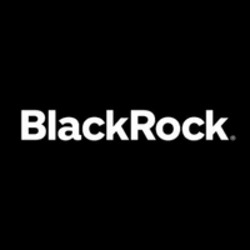}\\\footnotesize BUIDL\end{tabular} & \begin{tabular}{@{}c@{}}\includegraphics[height=4ex]{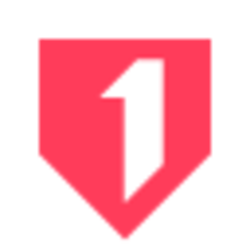}\\\footnotesize C1USD\end{tabular} & \begin{tabular}{@{}c@{}}\includegraphics[height=4ex]{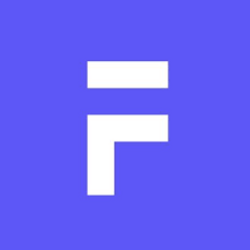}\\\footnotesize FIGR\_HELOC\end{tabular} & \begin{tabular}{@{}c@{}}\includegraphics[height=4ex]{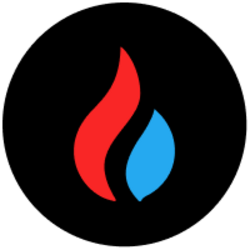}\\HTX\end{tabular} \\
\begin{tabular}{@{}c@{}}\includegraphics[height=4ex]{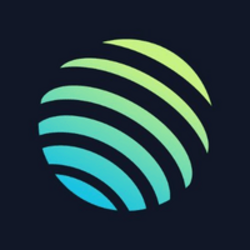}\\JLP\end{tabular} & \begin{tabular}{@{}c@{}}\includegraphics[height=4ex]{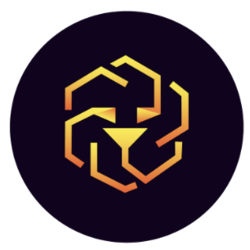}\\LEO\end{tabular} & \begin{tabular}{@{}c@{}}\includegraphics[height=4ex]{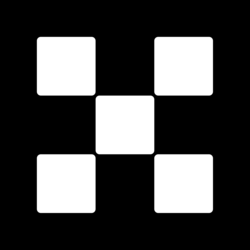}\\OKB\end{tabular} & \begin{tabular}{@{}c@{}}\includegraphics[height=4ex]{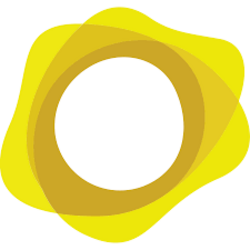}\\PAXG\end{tabular} & \begin{tabular}{@{}c@{}}\includegraphics[height=4ex]{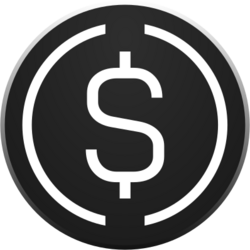}\\USDE\end{tabular} & \begin{tabular}{@{}c@{}}\includegraphics[height=4ex]{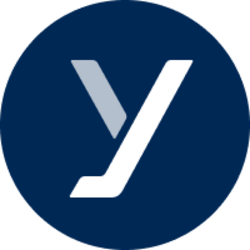}\\USYC\end{tabular} & \begin{tabular}{@{}c@{}}\includegraphics[height=4ex]{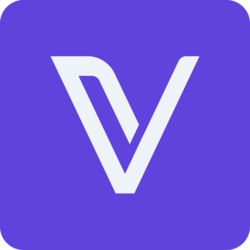}\\VET\end{tabular} \\
\begin{tabular}{@{}c@{}}\includegraphics[height=4ex]{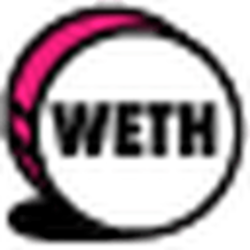}\\WETH\end{tabular} & \begin{tabular}{@{}c@{}}\includegraphics[height=4ex]{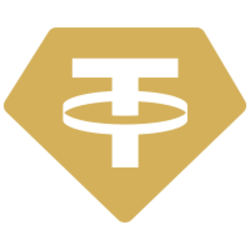}\\XAUT\end{tabular} & \begin{tabular}{@{}c@{}}\includegraphics[height=4ex]{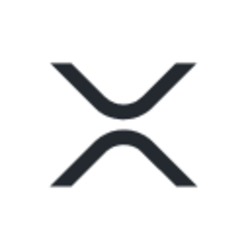}\\XRP\end{tabular} & \begin{tabular}{@{}c@{}}\includegraphics[height=4ex]{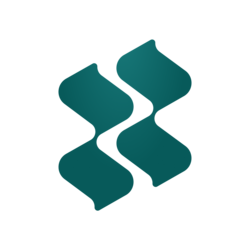}\\\footnotesize rsETH\end{tabular} &  &  &  \\
\end{tabular} \\[0.5em]
pass through certificate & \begin{tabular}{@{}ccccccc@{}}
\begin{tabular}{@{}c@{}}\includegraphics[height=4ex]{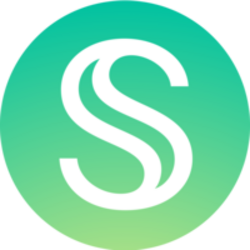}\\\footnotesize SUSDS\end{tabular} & \begin{tabular}{@{}c@{}}\includegraphics[height=4ex]{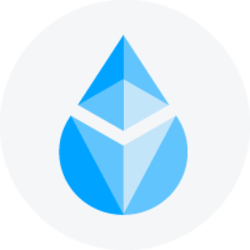}\\\footnotesize stETH\end{tabular} &  &  &  &  &  \\
\end{tabular} \\[0.5em]
payment-in-kind & \begin{tabular}{@{}ccccccc@{}}
\begin{tabular}{@{}c@{}}\includegraphics[height=4ex]{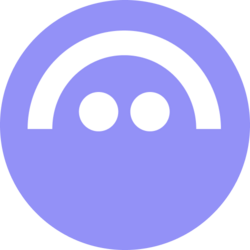}\\AAVE\end{tabular} & \begin{tabular}{@{}c@{}}\includegraphics[height=4ex]{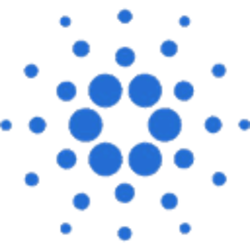}\\ADA\end{tabular} & \begin{tabular}{@{}c@{}}\includegraphics[height=4ex]{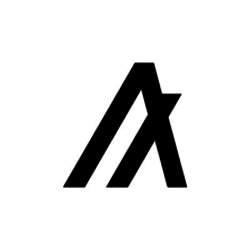}\\ALGO\end{tabular} & \begin{tabular}{@{}c@{}}\includegraphics[height=4ex]{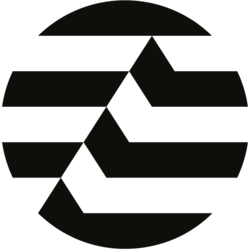}\\APT\end{tabular} & \begin{tabular}{@{}c@{}}\includegraphics[height=4ex]{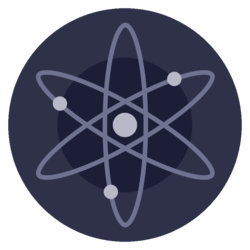}\\ATOM\end{tabular} & \begin{tabular}{@{}c@{}}\includegraphics[height=4ex]{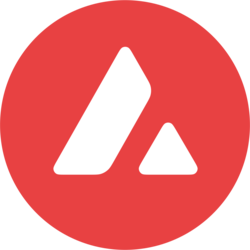}\\AVAX\end{tabular} & \begin{tabular}{@{}c@{}}\includegraphics[height=4ex]{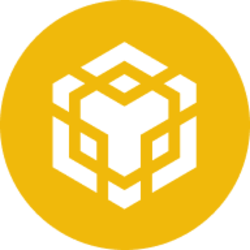}\\BNB\end{tabular} \\
\begin{tabular}{@{}c@{}}\includegraphics[height=4ex]{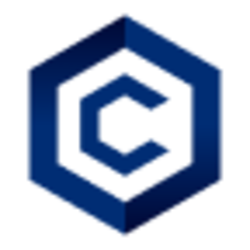}\\CRO\end{tabular} & \begin{tabular}{@{}c@{}}\includegraphics[height=4ex]{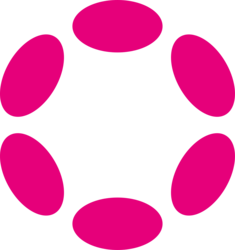}\\DOT\end{tabular} & \begin{tabular}{@{}c@{}}\includegraphics[height=4ex]{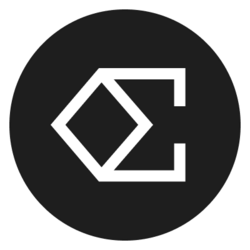}\\ENA\end{tabular} & \begin{tabular}{@{}c@{}}\includegraphics[height=4ex]{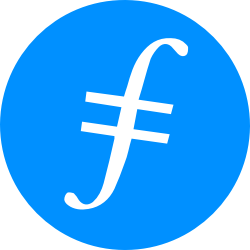}\\FIL\end{tabular} & \begin{tabular}{@{}c@{}}\includegraphics[height=4ex]{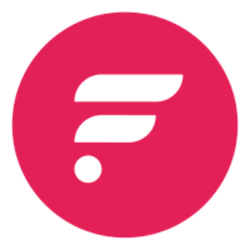}\\FLR\end{tabular} & \begin{tabular}{@{}c@{}}\includegraphics[height=4ex]{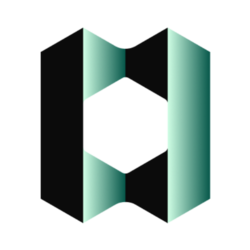}\\HASH\end{tabular} & \begin{tabular}{@{}c@{}}\includegraphics[height=4ex]{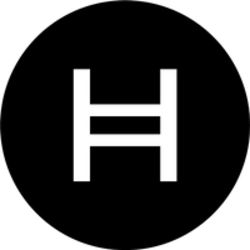}\\HBAR\end{tabular} \\
\begin{tabular}{@{}c@{}}\includegraphics[height=4ex]{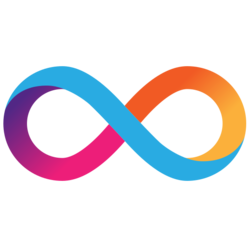}\\ICP\end{tabular} & \begin{tabular}{@{}c@{}}\includegraphics[height=4ex]{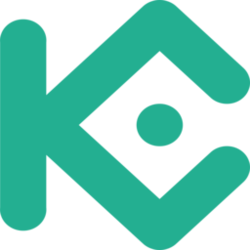}\\KCS\end{tabular} & \begin{tabular}{@{}c@{}}\includegraphics[height=4ex]{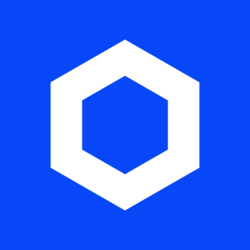}\\LINK\end{tabular} & \begin{tabular}{@{}c@{}}\includegraphics[height=4ex]{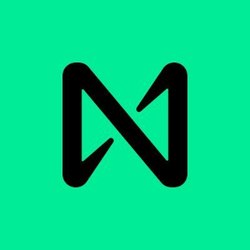}\\NEAR\end{tabular} & \begin{tabular}{@{}c@{}}\includegraphics[height=4ex]{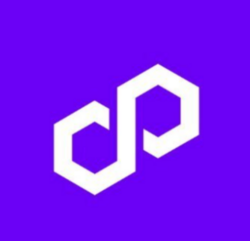}\\POL\end{tabular} & \begin{tabular}{@{}c@{}}\includegraphics[height=4ex]{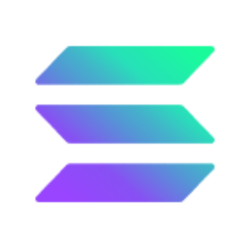}\\SOL\end{tabular} & \begin{tabular}{@{}c@{}}\includegraphics[height=4ex]{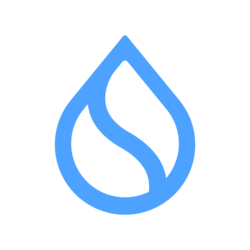}\\SUI\end{tabular} \\
\begin{tabular}{@{}c@{}}\includegraphics[height=4ex]{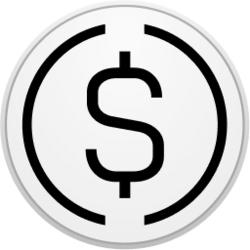}\\\footnotesize SUSDE\end{tabular} & \begin{tabular}{@{}c@{}}\includegraphics[height=4ex]{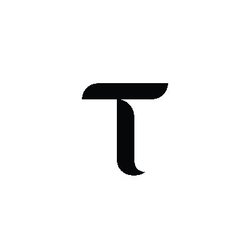}\\TAO\end{tabular} & \begin{tabular}{@{}c@{}}\includegraphics[height=4ex]{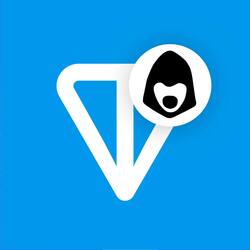}\\TON\end{tabular} & \begin{tabular}{@{}c@{}}\includegraphics[height=4ex]{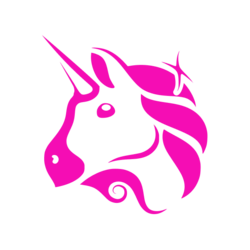}\\UNI\end{tabular} & \begin{tabular}{@{}c@{}}\includegraphics[height=4ex]{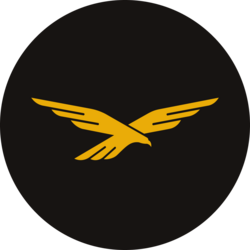}\\WLFI\end{tabular} &  &  \\
\end{tabular} \\[0.5em]
voting equity & \begin{tabular}{@{}ccccccc@{}}
\begin{tabular}{@{}c@{}}\includegraphics[height=4ex]{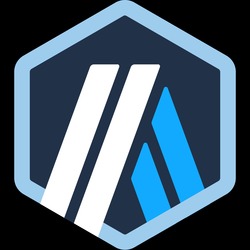}\\ARB\end{tabular} & \begin{tabular}{@{}c@{}}\includegraphics[height=4ex]{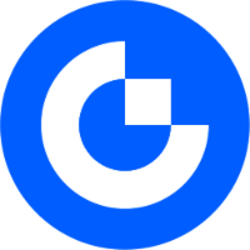}\\GT\end{tabular} & \begin{tabular}{@{}c@{}}\includegraphics[height=4ex]{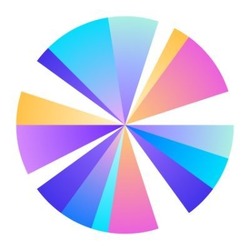}\\SKY\end{tabular} &  &  &  &  \\
\end{tabular} \\[0.5em]
\end{longtable}

%% file: tables/centralisation_table.tex
\begin{longtable}{p{0.20\textwidth} p{0.80\textwidth}}
\caption{Centralisation classification for the top 100 assets}\label{tab: centralisation}\\
\toprule
Centralisation & Assets \\
\midrule
\endfirsthead

\toprule
Centralisation & Assets \\
\midrule
\endhead

\bottomrule
\endfoot

\bottomrule
\endlastfoot

centralised & \begin{tabular}{@{}ccccccc@{}}
\begin{tabular}{@{}c@{}}\includegraphics[height=4ex]{figures/logos/bfusd.png}\\\footnotesize BFUSD\end{tabular} & \begin{tabular}{@{}c@{}}\includegraphics[height=4ex]{figures/logos/bgb.png}\\BGB\end{tabular} & \begin{tabular}{@{}c@{}}\includegraphics[height=4ex]{figures/logos/bnsol.png}\\\footnotesize BNSOL\end{tabular} & \begin{tabular}{@{}c@{}}\includegraphics[height=4ex]{figures/logos/bsc-usd.png}\\\footnotesize BSC-USD\end{tabular} & \begin{tabular}{@{}c@{}}\includegraphics[height=4ex]{figures/logos/buidl.png}\\\footnotesize BUIDL\end{tabular} & \begin{tabular}{@{}c@{}}\includegraphics[height=4ex]{figures/logos/c1usd.png}\\\footnotesize C1USD\end{tabular} & \begin{tabular}{@{}c@{}}\includegraphics[height=4ex]{figures/logos/cc.png}\\CC\end{tabular} \\
\begin{tabular}{@{}c@{}}\includegraphics[height=4ex]{figures/logos/figr_heloc.png}\\\footnotesize FIGR\_HELOC\end{tabular} & \begin{tabular}{@{}c@{}}\includegraphics[height=4ex]{figures/logos/gt.png}\\GT\end{tabular} & \begin{tabular}{@{}c@{}}\includegraphics[height=4ex]{figures/logos/htx.png}\\HTX\end{tabular} & \begin{tabular}{@{}c@{}}\includegraphics[height=4ex]{figures/logos/jlp.png}\\JLP\end{tabular} & \begin{tabular}{@{}c@{}}\includegraphics[height=4ex]{figures/logos/kcs.png}\\KCS\end{tabular} & \begin{tabular}{@{}c@{}}\includegraphics[height=4ex]{figures/logos/lbtc.png}\\LBTC\end{tabular} & \begin{tabular}{@{}c@{}}\includegraphics[height=4ex]{figures/logos/leo.png}\\LEO\end{tabular} \\
\begin{tabular}{@{}c@{}}\includegraphics[height=4ex]{figures/logos/m.png}\\M\end{tabular} & \begin{tabular}{@{}c@{}}\includegraphics[height=4ex]{figures/logos/okb.png}\\OKB\end{tabular} & \begin{tabular}{@{}c@{}}\includegraphics[height=4ex]{figures/logos/paxg.png}\\PAXG\end{tabular} & \begin{tabular}{@{}c@{}}\includegraphics[height=4ex]{figures/logos/pi.png}\\PI\end{tabular} & \begin{tabular}{@{}c@{}}\includegraphics[height=4ex]{figures/logos/pyusd.png}\\\footnotesize PYUSD\end{tabular} & \begin{tabular}{@{}c@{}}\includegraphics[height=4ex]{figures/logos/rlusd.png}\\\footnotesize RLUSD\end{tabular} & \begin{tabular}{@{}c@{}}\includegraphics[height=4ex]{figures/logos/syrupusdt.png}\\\footnotesize SYRUPUSDT\end{tabular} \\
\begin{tabular}{@{}c@{}}\includegraphics[height=4ex]{figures/logos/trump.png}\\\footnotesize TRUMP\end{tabular} & \begin{tabular}{@{}c@{}}\includegraphics[height=4ex]{figures/logos/usd1.png}\\USD1\end{tabular} & \begin{tabular}{@{}c@{}}\includegraphics[height=4ex]{figures/logos/usdc.png}\\USDC\end{tabular} & \begin{tabular}{@{}c@{}}\includegraphics[height=4ex]{figures/logos/usdf.png}\\USDF\end{tabular} & \begin{tabular}{@{}c@{}}\includegraphics[height=4ex]{figures/logos/usdg.png}\\USDG\end{tabular} & \begin{tabular}{@{}c@{}}\includegraphics[height=4ex]{figures/logos/usds.png}\\USDS\end{tabular} & \begin{tabular}{@{}c@{}}\includegraphics[height=4ex]{figures/logos/usdt.png}\\USDT\end{tabular} \\
\begin{tabular}{@{}c@{}}\includegraphics[height=4ex]{figures/logos/usdt0.png}\\\footnotesize USDT0\end{tabular} & \begin{tabular}{@{}c@{}}\includegraphics[height=4ex]{figures/logos/usdtb.png}\\\footnotesize USDTB\end{tabular} & \begin{tabular}{@{}c@{}}\includegraphics[height=4ex]{figures/logos/usyc.png}\\USYC\end{tabular} & \begin{tabular}{@{}c@{}}\includegraphics[height=4ex]{figures/logos/wbeth.png}\\\footnotesize WBETH\end{tabular} & \begin{tabular}{@{}c@{}}\includegraphics[height=4ex]{figures/logos/wbt.png}\\WBT\end{tabular} & \begin{tabular}{@{}c@{}}\includegraphics[height=4ex]{figures/logos/wld.png}\\WLD\end{tabular} & \begin{tabular}{@{}c@{}}\includegraphics[height=4ex]{figures/logos/wlfi.png}\\WLFI\end{tabular} \\
\begin{tabular}{@{}c@{}}\includegraphics[height=4ex]{figures/logos/xaut.png}\\XAUT\end{tabular} & \begin{tabular}{@{}c@{}}\includegraphics[height=4ex]{figures/logos/xrp.png}\\XRP\end{tabular} & \begin{tabular}{@{}c@{}}\includegraphics[height=4ex]{figures/logos/cbbtc.png}\\\footnotesize cbBTC\end{tabular} & \begin{tabular}{@{}c@{}}\includegraphics[height=4ex]{figures/logos/rseth.png}\\\footnotesize rsETH\end{tabular} &  &  &  \\
\end{tabular} \\[0.5em]
decentralised & \begin{tabular}{@{}ccccccc@{}}
\begin{tabular}{@{}c@{}}\includegraphics[height=4ex]{figures/logos/aave.png}\\AAVE\end{tabular} & \begin{tabular}{@{}c@{}}\includegraphics[height=4ex]{figures/logos/algo.png}\\ALGO\end{tabular} & \begin{tabular}{@{}c@{}}\includegraphics[height=4ex]{figures/logos/apt.png}\\APT\end{tabular} & \begin{tabular}{@{}c@{}}\includegraphics[height=4ex]{figures/logos/aster.png}\\\footnotesize ASTER\end{tabular} & \begin{tabular}{@{}c@{}}\includegraphics[height=4ex]{figures/logos/atom.png}\\ATOM\end{tabular} & \begin{tabular}{@{}c@{}}\includegraphics[height=4ex]{figures/logos/avax.png}\\AVAX\end{tabular} & \begin{tabular}{@{}c@{}}\includegraphics[height=4ex]{figures/logos/bch.png}\\BCH\end{tabular} \\
\begin{tabular}{@{}c@{}}\includegraphics[height=4ex]{figures/logos/bnb.png}\\BNB\end{tabular} & \begin{tabular}{@{}c@{}}\includegraphics[height=4ex]{figures/logos/btc.png}\\BTC\end{tabular} & \begin{tabular}{@{}c@{}}\includegraphics[height=4ex]{figures/logos/cro.png}\\CRO\end{tabular} & \begin{tabular}{@{}c@{}}\includegraphics[height=4ex]{figures/logos/dai.png}\\DAI\end{tabular} & \begin{tabular}{@{}c@{}}\includegraphics[height=4ex]{figures/logos/ena.png}\\ENA\end{tabular} & \begin{tabular}{@{}c@{}}\includegraphics[height=4ex]{figures/logos/etc.png}\\ETC\end{tabular} & \begin{tabular}{@{}c@{}}\includegraphics[height=4ex]{figures/logos/eth.png}\\ETH\end{tabular} \\
\begin{tabular}{@{}c@{}}\includegraphics[height=4ex]{figures/logos/fbtc.png}\\FBTC\end{tabular} & \begin{tabular}{@{}c@{}}\includegraphics[height=4ex]{figures/logos/flr.png}\\FLR\end{tabular} & \begin{tabular}{@{}c@{}}\includegraphics[height=4ex]{figures/logos/hbarx.png}\\\footnotesize HBARX\end{tabular} & \begin{tabular}{@{}c@{}}\includegraphics[height=4ex]{figures/logos/jitosol.png}\\\footnotesize JitoSOL\end{tabular} & \begin{tabular}{@{}c@{}}\includegraphics[height=4ex]{figures/logos/kas.png}\\KAS\end{tabular} & \begin{tabular}{@{}c@{}}\includegraphics[height=4ex]{figures/logos/khype.png}\\\footnotesize KHYPE\end{tabular} & \begin{tabular}{@{}c@{}}\includegraphics[height=4ex]{figures/logos/link.png}\\LINK\end{tabular} \\
\begin{tabular}{@{}c@{}}\includegraphics[height=4ex]{figures/logos/ltc.png}\\LTC\end{tabular} & \begin{tabular}{@{}c@{}}\includegraphics[height=4ex]{figures/logos/lseth.png}\\\footnotesize LsETH\end{tabular} & \begin{tabular}{@{}c@{}}\includegraphics[height=4ex]{figures/logos/mnt.png}\\MNT\end{tabular} & \begin{tabular}{@{}c@{}}\includegraphics[height=4ex]{figures/logos/near.png}\\NEAR\end{tabular} & \begin{tabular}{@{}c@{}}\includegraphics[height=4ex]{figures/logos/ondo.png}\\ONDO\end{tabular} & \begin{tabular}{@{}c@{}}\includegraphics[height=4ex]{figures/logos/pepe.png}\\PEPE\end{tabular} & \begin{tabular}{@{}c@{}}\includegraphics[height=4ex]{figures/logos/pol.png}\\POL\end{tabular} \\
\begin{tabular}{@{}c@{}}\includegraphics[height=4ex]{figures/logos/pump.png}\\PUMP\end{tabular} & \begin{tabular}{@{}c@{}}\includegraphics[height=4ex]{figures/logos/shib.png}\\SHIB\end{tabular} & \begin{tabular}{@{}c@{}}\includegraphics[height=4ex]{figures/logos/sky.png}\\SKY\end{tabular} & \begin{tabular}{@{}c@{}}\includegraphics[height=4ex]{figures/logos/sol.png}\\SOL\end{tabular} & \begin{tabular}{@{}c@{}}\includegraphics[height=4ex]{figures/logos/sui.png}\\SUI\end{tabular} & \begin{tabular}{@{}c@{}}\includegraphics[height=4ex]{figures/logos/susde.png}\\\footnotesize SUSDE\end{tabular} & \begin{tabular}{@{}c@{}}\includegraphics[height=4ex]{figures/logos/tao.png}\\TAO\end{tabular} \\
\begin{tabular}{@{}c@{}}\includegraphics[height=4ex]{figures/logos/ton.png}\\TON\end{tabular} & \begin{tabular}{@{}c@{}}\includegraphics[height=4ex]{figures/logos/trx.png}\\TRX\end{tabular} & \begin{tabular}{@{}c@{}}\includegraphics[height=4ex]{figures/logos/uni.png}\\UNI\end{tabular} & \begin{tabular}{@{}c@{}}\includegraphics[height=4ex]{figures/logos/usde.png}\\USDE\end{tabular} & \begin{tabular}{@{}c@{}}\includegraphics[height=4ex]{figures/logos/wbnb.png}\\WBNB\end{tabular} & \begin{tabular}{@{}c@{}}\includegraphics[height=4ex]{figures/logos/weth.png}\\WETH\end{tabular} & \begin{tabular}{@{}c@{}}\includegraphics[height=4ex]{figures/logos/wsteth.png}\\\footnotesize WSTETH\end{tabular} \\
\begin{tabular}{@{}c@{}}\includegraphics[height=4ex]{figures/logos/xlm.png}\\XLM\end{tabular} & \begin{tabular}{@{}c@{}}\includegraphics[height=4ex]{figures/logos/xmr.png}\\XMR\end{tabular} & \begin{tabular}{@{}c@{}}\includegraphics[height=4ex]{figures/logos/zec.png}\\ZEC\end{tabular} & \begin{tabular}{@{}c@{}}\includegraphics[height=4ex]{figures/logos/reth.png}\\rETH\end{tabular} & \begin{tabular}{@{}c@{}}\includegraphics[height=4ex]{figures/logos/steth.png}\\\footnotesize stETH\end{tabular} &  &  \\
\end{tabular} \\[0.5em]
hybrid & \begin{tabular}{@{}ccccccc@{}}
\begin{tabular}{@{}c@{}}\includegraphics[height=4ex]{figures/logos/ada.png}\\ADA\end{tabular} & \begin{tabular}{@{}c@{}}\includegraphics[height=4ex]{figures/logos/arb.png}\\ARB\end{tabular} & \begin{tabular}{@{}c@{}}\includegraphics[height=4ex]{figures/logos/doge.png}\\DOGE\end{tabular} & \begin{tabular}{@{}c@{}}\includegraphics[height=4ex]{figures/logos/dot.png}\\DOT\end{tabular} & \begin{tabular}{@{}c@{}}\includegraphics[height=4ex]{figures/logos/fil.png}\\FIL\end{tabular} & \begin{tabular}{@{}c@{}}\includegraphics[height=4ex]{figures/logos/hash.png}\\HASH\end{tabular} & \begin{tabular}{@{}c@{}}\includegraphics[height=4ex]{figures/logos/hbar.png}\\HBAR\end{tabular} \\
\begin{tabular}{@{}c@{}}\includegraphics[height=4ex]{figures/logos/hype.png}\\HYPE\end{tabular} & \begin{tabular}{@{}c@{}}\includegraphics[height=4ex]{figures/logos/icp.png}\\ICP\end{tabular} & \begin{tabular}{@{}c@{}}\includegraphics[height=4ex]{figures/logos/qnt.png}\\QNT\end{tabular} & \begin{tabular}{@{}c@{}}\includegraphics[height=4ex]{figures/logos/susds.png}\\\footnotesize SUSDS\end{tabular} & \begin{tabular}{@{}c@{}}\includegraphics[height=4ex]{figures/logos/syrupusdc.png}\\\footnotesize SYRUPUSDC\end{tabular} & \begin{tabular}{@{}c@{}}\includegraphics[height=4ex]{figures/logos/vet.png}\\VET\end{tabular} & \begin{tabular}{@{}c@{}}\includegraphics[height=4ex]{figures/logos/wbtc.png}\\WBTC\end{tabular} \\
\begin{tabular}{@{}c@{}}\includegraphics[height=4ex]{figures/logos/weeth.png}\\\footnotesize WEETH\end{tabular} &  &  &  &  &  &  \\
\end{tabular} \\[0.5em]
\end{longtable}

%% file: sections/case_studies.tex
\section{Case studies, \ac{tradfi} analogues and regulatory implications}
A token’s classification determines not only the economic model and attributes that merit analysis, but also the regulatory framework and prudential obligations that apply. Each dimension of the taxonomy corresponds to a legal or supervisory risk category: For example, the technical standard and base chain affect interoperability and operational resilience relevant for assessing systemic and market-infrastructure risk. The reference asset defines the core financial exposure, guiding whether the token function as a payment instrument, asset reference token or investment product. Minting yield and redemption mechanisms determine the existence of issuer liabilities and potential investor- or consumer-protection triggers.

Where a meaningful parallel exists, these links imply clear decision rules for assigning tokens to regulatory \enquote{buckets}, let us map some tokens onto well-documented traditional instruments, and by analogy offer reference points for discussing possible suitable regulatory treatments, and for handling edge cases consistently across contexts.

In the case studies that follow, we apply this feature-based lens to show how distinct crypto designs can be mapped to well-documented \ac{tradfi} instruments (e.g., secured funding, depositary receipts, pass-throughs). For each case study, we list the taxonomy instance built with explicit feature inputs (see design of taxonomy dataclass instances, type and examples of feature inputs, and decision trees in \autoref{sec:dataclass_design}), the derived features, and analogy to \ac{tradfi} instruments. For the ownership makeup subdimension of the centralisation metric, the results are currently based solely on Ethereum-based DEX activity; a more detailed and comprehensive categorisation would require extending the analysis to additional base chains. The analogies are introduced within individual case studies to illustrate how specific crypto features resemble established financial instruments. They serve to translate regulatory intuition into familiar frameworks. More cases for BTC, UNI, DAI, HBAR, cbETH, and XRP can be found in \autoref{app: case study}.

\begin{CaseBox}[Case: WBTC — depositary receipt]
\textbf{Taxonomy instance (explicit feature inputs)}\\[-0.25em]
\begin{lstlisting}[language=json, escapeinside={(*@}{@*)}]
technical standard: "erc-20";
function: "security";
dic_critical_resource(*@\footnote{The WBTC DAO multisig have 13 signers with eight needed to achieve quorum on any matter. See \url{https://www.theblock.co/post/189991/wrapped-bitcoin-dao-removes-ftx-nine-others-in-move-to-new-multisig?utm_source=chatgpt.com} and \url{https://www.theblock.co/post/189991/wrapped-bitcoin-dao-removes-ftx-nine-others-in-move-to-new-multisig} }@*)(*@\footnote{WBTC is minted 1:1 against BTC that BitGo holds in cold storage, BitGo holds custodian control over minting and redemption.}@*): {
    "gov_rule_change": {"upgrade_authorities": 8},
    "mint_authority": {"custodial_issuers": 1},
    "red_reserve": {"reserve_custodians": 1}
};
minting type: "wrapped";
yield source: None;
distribution mechanism: None;
reference: "BTC";
is stablecoin: False;
redemption mechanism: "bridge_burn-and-release";
legal classification: "security";
form of claim: "in personam, against reserve, proceeds to issuers"
\end{lstlisting}

\vspace{0.5em}
\hrule
\vspace{0.5em}

\textbf{Properties (derived features)}\\[-0.25em] 
\begin{lstlisting}[language=json]
centralisation: "hybrid",
tradfi_analogy: "repositary_receipt"
\end{lstlisting}
\vspace{0.5em}
\hrule
\vspace{0.5em}

WBTC is an ERC-20 token issued 1:1 against BTC held with a centralised custodian (BitGo), with mint routed through approved merchants, thus categorised as hybrid in centralisation dimension. It mobilizes idle BTC for cross-chain use, such as collateral, liquidity, and settlement. Governance and operations are coordinated by an industry consortium, and supply is transparent on-chain. WBTC represents an off-chain reserve held in custody, relies on standardised issuance and cancellation workflows, and offers a redemption path that anchors price to the underlying. 

The setup, asset referenced redemption, and non-yield bearing, echo the architecture in Depositary Receipt (DR) programs\footnote{\ac{mifid} II defines \emph{depositary receipts} as: \enquote{those securities which are negotiable on the capital market and which represent ownership of the securities of a non-domiciled issuer while being able to be admitted to trading on a regulated market and traded independently of the securities of the non-domiciled issuer;} Directive 2014/65/EU (MiFID II), Article 4(1)(45). See \url{https://www.esma.europa.eu/publications-and-data/interactive-single-rulebook/mifid-ii/article-4-definitions}.} Economic exposure is intentionally 1 : 1 to the referenced asset. However, DRs convey shareholder/dividend rights via regulated securities infrastructure, WBTC does not. Holders only access a redemption pathway mediated by merchants/custodians. 
\end{CaseBox}

\begin{CaseBox}[Case: stETH — \ac{ptc}]
\textbf{Taxonomy instance (explicit feature inputs)}\\[-0.25em]
\begin{lstlisting}[language=json, escapeinside={(*@}{@*)}]
technical standard: "erc-20";
function: "security";
dic_critical_resource: {
    "gov_voting":{"quorum_threshold": 3(*@\footnote{Lido has different thresholds (at least 3) for ops multsig for special cases. See \url{https://research.lido.fi/t/lido-dao-ops-multisigs-policy/4115}}@*)},
    "yield_reward_policy": {"reward_rates": 5(*@\footnote{Current stETH Oracle set consists of 9 participants with a quorum of 5, see \url{https://docs.lido.fi/guides/oracle-operator-manual}.}@*)},
    "red_mechanism": {"gatekeepers_whitelisting": 3(*@\footnote{GateSeal is a contract that allows the designated account to instantly put a set of contracts on pause, GateSeal committee is a 3/6 multisig, see \url{https://research.lido.fi/t/lido-dual-governance-explainer-research-distillation/7132}}@*)}
};
minting type: "staking";
yield source: "staking_rewards";
distribution mechanism: "quantity_accrual";
reference: ETH;
is stablecoin: False;
redemption mechanism: "queued_withdrawal";
legal classification: "other crypto-asset";
form of claim: "in personam, against reserve, proceeds to holders"
\end{lstlisting}
\vspace{0.5em}
\hrule
\vspace{0.5em}

\textbf{Properties (derived features)}\\[-0.25em] 
\begin{lstlisting}[language=json]
centralisation: "decentralised",
tradfi_analogy: "pass-through certificates".
\end{lstlisting}
\vspace{0.5em}
\hrule
\vspace{0.5em}

stETH, a type of \ac{lst}s, represent staked positions on a proof-of-stake network, tokenizing the right to withdraw or trade the underlying staked asset (redemption to ETH) plus any accrued rewards (staking\_rewards yield type). The token’s value tracks both the principal and staking yield, allowing holders to retain liquidity while participating in network consensus. \ac{lst}s are legally treated as other crypto-assets by US SEC because they do not involve the offer and sale of securities, and therefore fail the Howey test\footnote{See SEC, Division of Corporation Finance, \enquote{Statement on Certain Liquid Staking Activities}, Aug. 5, 2025}.

\ac{lst}s pass through staking rewards from validator operations to token holders, which is analogous to \ac{ptc} under US Reg AB / Reg AB II regulation \footnote{PTCs are a form of asset-backed security under the SEC’s ABS framework: U.S. Securities and Commission, \enquote{some ABS transactions consist of simple pass-through certificates representing a pro rata share of the cash flows from the underlying asset pool}. See \url{https://www.federalregister.gov/documents/2005/01/07/05-53/asset-backed-securities}}, defined as securities that transmit interest and principal from a pool of assets to investors. Both distribute income directly from an underlying productive pool, and redemption reflects the performance of that pool rather than an issuer’s promise. 
\footnote{SEC Investor.gov: \enquote{The most basic types are pass-through participation certificates, which entitle the holder to a pro-rata share of all principal and interest payments made on the pool of mortgage loans.} See \url{https://www.investor.gov/introduction-investing/investing-basics/glossary/mortgage-backed-securities-and-collateralized}.} 

While EU law does not provide a security that grants investors an automatic, undivided pro-rata claim to all cash flows of the underlying assets, securitisations under the Securitisation Regulation (Reg. (EU) 2017/2402) entitle holders only to the cash flows allocated to their tranche by the transaction’s priority of payments(waterfall), subject to fees, reserves, and hedges.\footnote{\ac{esma}, SECR Article 2 (Definitions), Interactive Single Rulebook, See \url{https://www.esma.europa.eu/publications-and-data/interactive-single-rulebook/secr/article-2-definitions}.}.
\end{CaseBox}

%% file: sections/conclusion_and_discussion.tex
\section{Conclusion and Discussion}

This paper proposed a structured taxonomy of crypto-assets grounded in clearly defined, orthogonal dimensions. Building on prior literature, regulatory frameworks, and empirical asset mapping, the taxonomy captures the heterogeneous design space of crypto-assets in a way that is both analytically rigorous and practically applicable. By aligning specific feature combinations with analogues in traditional finance, the taxonomy contributes to improved regulatory clarity and investor understanding.

Representative case studies illustrate how taxonomy disentangles features for assets. The framework was operationalised through programmable data structure and validated against top 100 assets by market capitalisation. The distribution of their features reveals concentration patterns, of their \ac{tradfi} analogues show edge cases that challenge current categorisations.

Several open challenges remain. First, the taxonomy is static, whereas the crypto-asset ecosystem evolves rapidly. Mechanisms for continuous updating or protocol self-classification may be needed. Second, the current set of \ac{tradfi} analogues is not exclusive, and more work is needed to support comprehensive and jurisdiction-sensitive mappings across regulatory contexts. Finally, further work is needed to integrate this taxonomy with risk-based regulatory frameworks and cross-jurisdictional classification regimes. Future research could extend this taxonomy to cover dynamic protocol behaviour. As new crypto-asset designs emerge with higher complexity and regulatory relevance, systematic classification will be foundation for a transparent and clear oversight.

%% file: sections/appendix.tex
\appendix
\renewcommand{\thesection}{\Alph{section}}

\section{Reference Taxonomy Figures}
This appendix shows structure of several existing taxonomies.
\begin{figure}[htbp]
  \centering
  \includegraphics[width=0.95\linewidth]{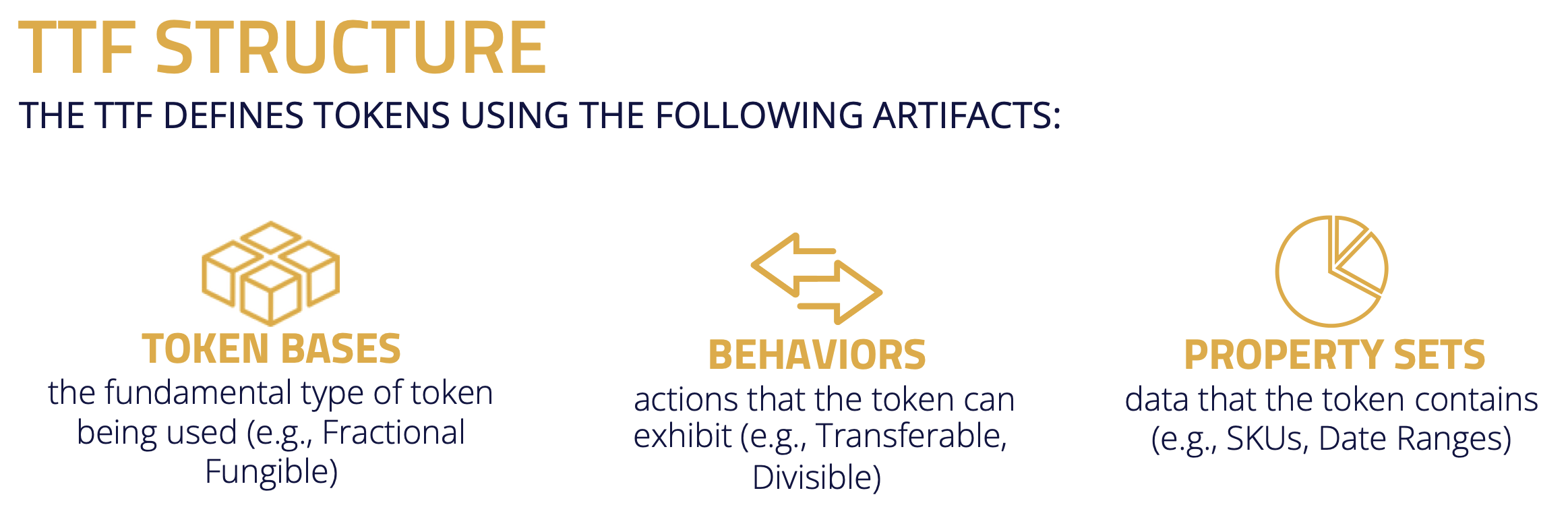}
  \caption{Token Taxonomy Framework (TTF).}
  \label{fig:tax-ttf-full}
\end{figure}

\begin{figure}[htbp]
  \centering
  \includegraphics[width=0.95\linewidth]{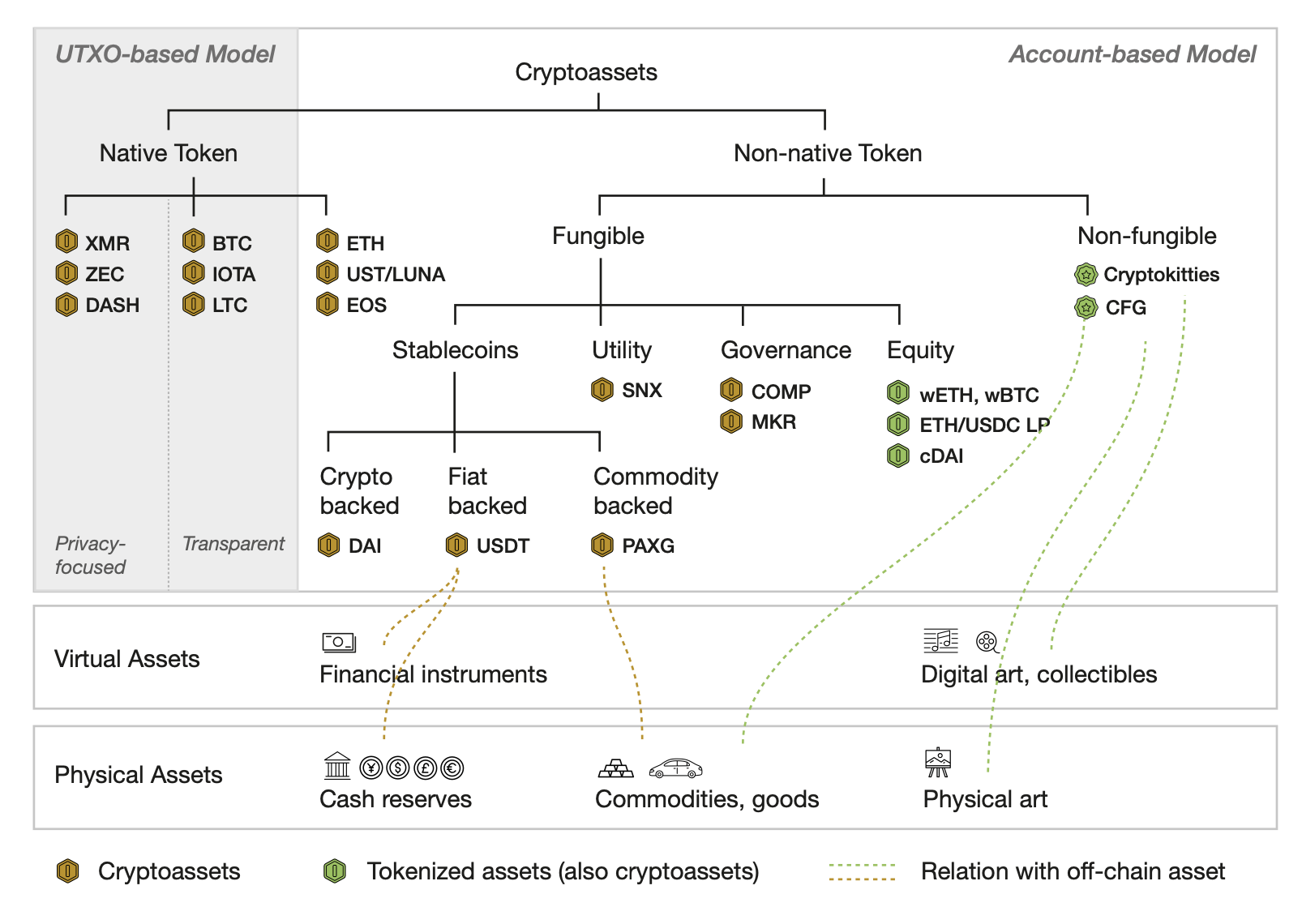}
  \caption{Token Taxonomy Framework (BIS).}
  \label{fig:tax-bis-full}
\end{figure}

\begin{figure}[htbp]
  \centering
  \includegraphics[width=0.95\linewidth]{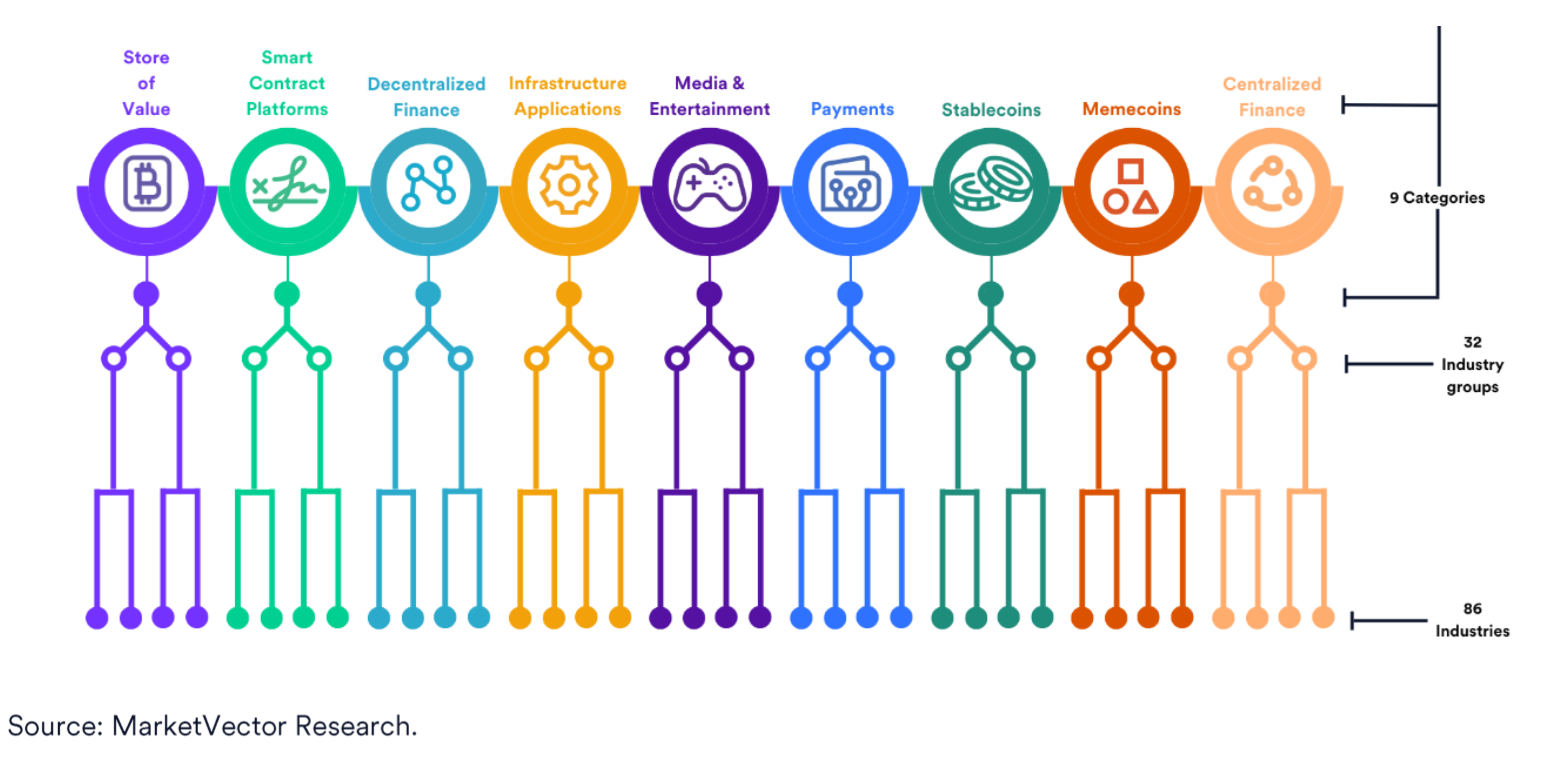}
  \caption{Token Taxonomy Framework (MarketVector).}
  \label{fig:tax-mv-full}
\end{figure}

\section{More case studies}
This appendix provides more case studies continuing \autoref{app: case study}.

\label{app: case study}
\begin{CaseBox}[Case: BTC — commodity]
\textbf{Taxonomy instance (explicit feature inputs)}\\[-0.25em]
\begin{lstlisting}[language=json, escapeinside={(*@}{@*)}]
technical standard: "native",
function: "utility",
dic_critical_resource(*@\footnote{Rule changes of Bitcoin is control by client maintainers and economic nodes. Bitcoin Core releases are produced and signed by multiple co-maintainers, see \url{https://bitcoin.org/en/about-us}, and each economic node independently enforces its chosen rule set, see \url{https://developer.bitcoin.org/devguide/block_chain.html}}@*): {"mint_authority": {"minting keys": 3}(*@\footnote{Number of parties controlling Bitcoin mining pools is current estimate from \url{https://chainspect.app/dashboard/decentralization}}@*)},
minting type: "consensus",
yield source: None
reference: None,
is stablecoin: False,
redemption mechanism: None,
legal classification: "other crypto-asset",
form of claim: "no claim"
\end{lstlisting}
\vspace{0.5em}
\hrule
\vspace{0.5em}

\textbf{Properties (derived features)}\\[-0.25em] 
\begin{lstlisting}[language=json]
centralisation: "decentralised"
tradfi_analogy: "commodity"
\end{lstlisting}
\vspace{0.5em}
\hrule
\vspace{0.5em}

Bitcoin (BTC) is the archetypal first-generation crypto asset, launched in 2009 as peer-to-peer electronic cash on a permissionless, proof-of-work network. In practice it plays three roles: a currency for transferring value with on-chain finality; a store of value (\enquote{digital gold}) given its predictable issuance; and a primary trading instrument that anchors liquidity across venues. 

For our taxonomy, BTC bears no native yield and confers no redemption right, it is best understood as a scarce, rivalrous digital good. Mining converts energy, hardware, and facilities into verifiable tokens while creating no corresponding liability, so newly minted BTC are non-financial assets that can be accumulated on balance sheets. As detailed in \cite{Rotta2022BitcoinCommodity}, BTC exhibits purchasing-power volatility, limited transactional acceptance, and persistent throughput constraints, features more consistent with a non-monetary, commodity-like asset than with a widely accepted medium of exchange.

\end{CaseBox}

\begin{CaseBox}[Case: UNI — voting equity share]
\textbf{Taxonomy instance (explicit feature inputs)}\\[-0.25em]
\begin{lstlisting}[language=json, escapeinside={(*@}{@*)}]
technical standard: "erc-20"; 
function: "governance"; 
dic_critical_resource(*@\footnote{The voting quorum and threshold are 40M and 1M UNIs, the number of parties in control of voting depends on the participation centralisation.}@*):  {};
minting type: "emission_governance"; 
yield source: "incentive_emissions"; 
distribution mechanism: "quantity\_accrual"; 
reference: None; 
is stablecoin: False; 
redemption mechanism: None; 
legal classification: "other crypto-asset";
form of claim: "no claim".
\end{lstlisting}
\vspace{0.5em}
\hrule
\vspace{0.5em}

\textbf{Properties (derived features)}\\[-0.25em] 
\begin{lstlisting}[language=json]
centralisation: "decentralised"
tradfi_analogy: "voting equity share"
\end{lstlisting}
\vspace{0.5em}
\hrule
\vspace{0.5em}
UNI is the governance token of the Uniswap protocol, granting holders the right to propose and vote on protocol parameters, treasury allocations, and upgrades. It bears no redemption right and typically no automatic yield. 

In functional terms, UNI most closely parallels a voting equity or membership share; holders influence governance outcomes and future policy decisions but lack legal ownership or claim on the protocol’s revenues. Distribution through airdrops and liquidity mining echoes equity grants aimed at decentralising control and rewarding early participation. Unlike corporate shares, UNI conveys no private-law rights or entitlement to profits; its governance power operates through smart contracts rather than shareholder resolutions. 
\end{CaseBox}

\pagebreak

\newpage

\begin{CaseBox}[Case: HBAR — payment in kind]
\textbf{Taxonomy instance (explicit feature inputs)}\\[-0.25em]
\begin{lstlisting}[language=json, escapeinside={(*@}{@*)}]
technical standard: "native";
function: "utility";
dic_critical_resource(*@\footnote{Explanation of dic\_critical\_resource inputs: The governance of HBAR is based on Hedera governance council, there are different voting threshold, no bounded number of parties, see \url{https://files.hedera.com/Hedera_COUNCIL-OVERVIEW_2022_OCT.pdf}.}@*): {
};
minting type: "pre_mined";
yield source: "staking_rewards";
distribution mechanism: "quantity_accrual";
reference: None;
is stablecoin: False;
legal classification: "other crypto-asset";
form of claim: "no claim"
\end{lstlisting}
\vspace{0.5em}
\hrule
\vspace{0.5em}

\textbf{Properties (derived features)}\\[-0.25em] 
\begin{lstlisting}[language=json]
centralisation: "decentralised"
tradfi_analogy: "payment in kind"
\end{lstlisting}
\vspace{0.5em}
\hrule
\vspace{0.5em}
HBAR is the native asset of the Hedera public network. Governance is led by a multi-industry Governing Council that steers network policy, authorizes releases from Treasury accounts holding the fixed, pre-minted 50B HBAR, and presently oversees a permissioned set of consensus nodes. HBAR has no reference asset and staking HBAR generates staking rewards without lock-ups or slashing. The yield is protocol-level and funded from a designated reward account, accruing over daily epochs and disbursed on user activity. 

HBAR’s staking rewards resemble a payment-in-kind (PIK) in economic effect: instead of periodic cash coupons, the holder’s position grows in kind as additional HBAR are credited to balance, so the claim increases over time without lock-up or cash distribution. As with PIK instruments, where interest is capitalised and added to principal, the unit count rises while liquidity remains available. \footnote{UK HMRC Corporate Finance Manual: PIK interest is satisfied by issuing further securities or capitalising interest into principal (see HMRC CFM11090 \url{https://www.gov.uk/hmrc-internal-manuals/corporate-finance-manual/cfm11090}).} 
\end{CaseBox}

\begin{CaseBox}[Case: DAI — repo]
\textbf{Taxonomy instance (explicit feature inputs)}\\[-0.25em]
\begin{lstlisting}[language=json, escapeinside={(*@}{@*)}]
technical standard: "erc-20";
functional: "utility";
dic_critical_resource(*@\footnote{The governance of DAI depends on MKR tokens, key parameters are set by holders' votes. No explicit threshold or quorum decided}@*): {
    "mint_data_param": {"oracle_aggregators": None(*@\footnote{DAI's oracle is median of Maker's trusted reference price, governance of MKR sets parameters to control Median's behaviour, see \url{https://docs.makerdao.com/smart-contract-modules/oracle-module/median-detailed-documentation, similarly the oracle for vault close decided (see below "redemption" domain)}}@*)},
},
minting type: "lock_and_mint";
yield source: None;
reference: "USD";
is stablecoin: True;
redemption mechanism: "protocal_par";
legal classification: "stable-value token";
form of claim: "no claim"
\end{lstlisting}
\vspace{0.5em}
\hrule
\vspace{0.5em}

\textbf{Properties (derived features)}\\[-0.25em] 
\begin{lstlisting}[language=json]
centralisation: "decentralised",
reference_asset: "asset-referenced tokens",
tradfi_analogy: "repo".
\end{lstlisting}
\vspace{0.5em}
\hrule
\vspace{0.5em}
DAI is a decentralised stablecoin issued by MakerDAO. Two issuance paths exist: (i) Vault minting. Users lock volatile collateral (e.g., ETH, wBTC) to borrow DAI against an over-collateralised position, repaying with a stability fee; and (ii) \ac{psm}. Users swap approved stablecoins (e.g., USDC) for DAI at or near par, and can swap back, expanding/contracting supply without liquidation risk. 

Once minted, all DAI are fungible. Holders can directly redeem USDC via Maker’s own contract. The PSM functions like a standing security financing\footnote{\ac{esma} SFTR Article3: "securities financing transaction means a repurchase transaction." See \url{https://www.esma.europa.eu/publications-and-data/interactive-single-rulebook/sftr/article-3-definitions}} (repo-style) facility\footnote{ECB: "repurchase operation is a liquidity-providing reverse transaction based on a repurchase agreement." See \url{https://www.ecb.europa.eu/services/glossary/html/glossr.en.html}}: users post stable collateral to obtain DAI at par and reverse the trade at will; governance-set fees (in/out) resemble funding spreads. 
\end{CaseBox}

\begin{CaseBox}[Case: cbETH — capitalising share class]
\textbf{Taxonomy instance (explicit feature inputs)}\\[-0.25em]
\begin{lstlisting}[language=json, escapeinside={(*@}{@*)}]
technical standard: "erc-20"; 
function: security; 
dic_critical_resource(*@\footnote{Coinbase is the staking provider and the issuer of cbETH, it plays roles of admin, owner, blacklister, masterMinter, minter, and pauser, hold custody for keys controlling the wallet addresses holding the staked ETH wrapped for cbETH, charges commission fee over underlying ETH, and credits staked ETH based on the published conversion rate when unwrapping.}@*): {
    "gov_rule_change": {"upgrade_authorites", 1},
    "mint_authority": {"custodial_issuers": 1},
    "yield_reward_policy": {"fee_levels": 1},
    "red_mechanism": {"settlement_custodians": 1},
};
minting type: "wrapped"; 
yield source: "staking_rewards"; 
distribution mechanism: "value_accrual"; 
reference: ETH;
is stablecoin: False;
redemption mechanism: "off-chain";
legal classification: "security or financial instrument";
form of claim: "in personam, against reserve, proceeds to holders".
\end{lstlisting}
\vspace{0.5em}
\hrule
\vspace{0.5em}

\textbf{Properties (derived features)}\\[-0.25em] 
\begin{lstlisting}[language=json]
centralisation: "centralised",
tradfi_analogy: "repo".
\end{lstlisting}
\vspace{0.5em}
\hrule
\vspace{0.5em}
cbETH, Coinbase Wrapped Staked ETH, is Coinbase’s liquid-staking wrapper for ETH. cbETH is minted and burned at a floating conversion rate to ETH that reflects underlying staked-ETH after rewards, penalties, and fees. Economically, holders keep a fixed unit count while the redemption value per unit accretes over time via the rising conversion rate. 

Functionally, cbETH resembles a \ac{esma}’s UCITS capitalising share-class, accreting instrument: instead of distributing coupons, returns are capitalised into the conversion rate, so value compounds until redemption or sale. Nevertheless, cbETH has no contractual coupon or maturity. 
\end{CaseBox}

\begin{CaseBox}[Case: XRP — commodity]
\textbf{Taxonomy instance (explicit feature inputs)}\\[-0.25em]
\begin{lstlisting}[language=json, escapeinside={(*@}{@*)}]
technical standard: "native",
dic critical resource: {
  "gov_voting": {"validator_curation": 1(*@\footnote{XRPL Foundation is the sole publisher of default UNL, see \url{https://xrpl.org/blog/2025/move-to-the-new-xrpl-foundation-commences}.}@*)
  },
  "mint_authority": {"custodial_issuers": 1(*@\footnote{Ripple’s in control of regular XRP on-ledger escrow releases, see \url{https://ripple.com/insights/explanation-ripples-xrp-escrow/}.}@*)},
};
function: "utility";
minting type: "pre_mined";
yield source: None;
reference: None;
is stablecoin: False;
redemption mechanism: None;
legal classification: "other crypto-asset"
form of claim: "no claim"
\end{lstlisting}
\vspace{0.5em}
\hrule
\vspace{0.5em}

\textbf{Properties (derived features)}\\[-0.25em] 
\begin{lstlisting}[language=json]
centralisation: "centralised"
tradfi_analogy: "commodity"
\end{lstlisting}
\vspace{0.5em}
\hrule
\vspace{0.5em}
XRP is the native asset of the XRP Ledger (XRPL), a public L1 optimised for fast settlement and built-in features such as on-ledger AMM and DEX. XRP has no reference asset and was pre mined at genesis (100B) with a long-running Ripple escrow program that releases tokens on a monthly cadence. Protocol changes follow XRPL’s \enquote{amendment} process: an amendment activates only after maintaining over 80\% support from trusted validators for two consecutive weeks; operators can also change the validator lists (UNLs) they trust, with the default list now published by the XRPL Foundation as part of a 2025 migration.

Similar to BTC, XRP bears no native yield and confers no redemption right; it is best understood as a scarce, rivalrous digital good. Unlike mined issuance, XRP’s entire supply (100 billion) was pre-minted and then distributed over time without creating a corresponding issuer liability; ownership is purely bearer-style control of units on the ledger. The protocol also destroys a small transaction fee with each transfer, reinforcing scarcity at the margin. In economic terms, XRP exhibits purchasing-power volatility and limited transactional acceptance.
\end{CaseBox}

\begin{CaseBox}[Case: HBARX — capotalising share class]
\textbf{Taxonomy instance (explicit feature inputs)}\\[-0.25em]
\begin{lstlisting}[language=json, escapeinside={(*@}{@*)}]
technical standard: "other",
dic critical resource: {
};
function: "other";
minting type: "staking";
yield source: "staking_rewards";
reference: HBAR;
is stablecoin: False;
redemption mechanism: "queued_withdrawal";
legal classification: "other crypto-asset"
form of claim: "no claim"
\end{lstlisting}
\vspace{0.5em}
\hrule
\vspace{0.5em}

\textbf{Properties (derived features)}\\[-0.25em] 
\begin{lstlisting}[language=json]
centralisation: "centralised"
tradfi_analogy: "commodity"
\end{lstlisting}
\vspace{0.5em}
\hrule
\vspace{0.5em}
HBARX is a liquid staking token issued by the Stader protocol on the Hedera network using the Hedera Token Service (HTS). It represents a proportional claim on a pooled HBAR staking position and allows holders to retain liquidity while earning native staking rewards.

HBARX is minted when users stake HBAR through Stader and is redeemable on-chain by burning HBARX to withdraw HBAR after an unbonding period. The token’s exchange rate to HBAR increases over time as staking rewards accrue. HBARX involves no custodial reserve, off-chain issuer redemption, or legal issuer obligation. 
\end{CaseBox}

%% file: references.bib
@techreport{Leinweber2024AWorld,
    title = {{A Classification Framework for Digital Assets Sorting out the crypto world}},
    year = {2024},
    author = {Leinweber, Martin and Digital, Jonas Weber and Analyst, Assets},
    institution = {MarketVector}
}

@article{Moin2019ADesigns,
    title = {{A Classification Framework for Stablecoin Designs}},
    year = {2019},
    journal = {Lecture Notes in Computer Science},
    author = {Moin, Amani and Sirer, Emin Gün and Sekniqi, Kevin},
    month = {9},
    pages = {174},
    url = {http://arxiv.org/abs/1910.10098},
    institution = {arXiv},
    arxivId = {1910.10098}
}

@article{Nickerson2013ASystems,
    title = {{A Method for Taxonomy Development and its Application in Information Systems}},
    year = {2013},
    journal = {European Journal of Information Systems},
    author = {Nickerson, Robert and Varshney, Upkar and Muntermann, Jan},
    month = {12},
    volume = {22},
    doi = {10.1057/ejis.2012.26}
}

@article{Puschmann2024AFinance,
    title = {{A Taxonomy for Decentralized Finance}},
    year = {2024},
    journal = {Journal of Financial Markets, Institutions and Money},
    author = {Puschmann, Thomas}
}

@misc{Tasca2019AClassification,
    title = {{A Taxonomy of Blockchain Technologies: Principles of Identification and Classification}},
    year = {2019},
    booktitle = {Ledger},
    author = {Tasca, Paolo and Tessone, Claudio J.},
    pages = {1--39},
    volume = {4},
    publisher = {University Library System, University of Pittsburgh},
    doi = {10.5195/LEDGER.2019.140},
    issn = {23795980},
    arxivId = {1708.04872}
}

@techreport{MilkenInstitute2023AAssets,
    title = {{A Taxonomy of Digital Assets}},
    year = {2023},
    author = {{Milken Institute}},
    institution = {Milken Institute},
    address = {Santa Monica, CA}
}

@techreport{CryptoCouncilforInnovation2025AccountingLSTs,
    title = {{Accounting Guidance Request for Liquid Staking Tokens (LSTs)}},
    year = {2025},
    author = {{Crypto Council for Innovation} and {Proof of Stake Alliance}},
    url = {https://cryptocouncil.org/},
    institution = {CCI and POSA}
}

@article{Rotta2022BitcoinCommodity,
    title = {{Bitcoin as a digital commodity}},
    year = {2022},
    journal = {New Political Economy},
    author = {Rotta, Tomás N. and Paran{\'{a}}, Edemilson},
    number = {6},
    pages = {1046--1061},
    volume = {27},
    publisher = {Routledge},
    doi = {10.1080/13563467.2022.2054966},
    issn = {14699923}
}

@misc{Abdelmaboud2022BlockchainDirections,
    title = {{Blockchain for IoT Applications: Taxonomy, Platforms, Recent Advances, Challenges and Future Research Directions}},
    year = {2022},
    booktitle = {Electronics (Switzerland)},
    author = {Abdelmaboud, Abdelzahir and Ahmed, Abdelmuttlib Ibrahim Abdalla and Abaker, Mohammed and Eisa, Taiseer Abdalla Elfadil and Albasheer, Hashim and Ghorashi, Sara Abdelwahab and Karim, Faten Khalid},
    number = {4},
    month = {2},
    volume = {11},
    publisher = {MDPI},
    doi = {10.3390/electronics11040630},
    issn = {20799292}
}

@article{Loporchio2024ComparingNetworks,
    title = {{Comparing Ethereum fungible and non-fungible tokens: an analysis of transfer networks}},
    year = {2024},
    journal = {Applied Network Science},
    author = {Loporchio, Matteo and Di Francesco Maesa, Damiano and Bernasconi, Anna and Ricci, Laura},
    number = {1},
    month = {12},
    volume = {9},
    publisher = {Springer Nature},
    doi = {10.1007/s41109-024-00682-8},
    issn = {23648228}
}

@article{Zetzsche2024CryptoCustody,
    title = {{Crypto custody}},
    year = {2024},
    journal = {Capital Markets Law Journal},
    author = {Zetzsche, Dirk and Sinnig, Julia and Nikolakopoulou, Areti},
    number = {3},
    month = {7},
    pages = {207--229},
    volume = {19},
    publisher = {Oxford University Press},
    doi = {10.1093/cmlj/kmae010},
    issn = {17507227}
}

@misc{Ito2024CryptoeconomicsOpinions,
    title = {{Cryptoeconomics and Tokenomics as Economics: A Survey with Opinions}},
    year = {2024},
    author = {Ito, Kensuke},
    month = {12},
    doi = {10.48550/arXiv.2407.15715}
}

@techreport{BlackVogel2025DARTEWashington,
    title = {{DARTE SERIES Washington}},
    year = {2025},
    author = {{BlackVogel}},
    institution = {BlackVogel}
}

@article{Sun2024DecentralizationPolls,
    title = {{Decentralization illusion in Decentralized Finance: Evidence from tokenized voting in MakerDAO polls}},
    year = {2024},
    journal = {Journal of Financial Stability},
    author = {Sun, Xiaotong and Stasinakis, Charalampos and Sermpinis, Georgios},
    month = {8},
    volume = {73},
    publisher = {Elsevier B.V.},
    doi = {10.1016/j.jfs.2024.101286},
    issn = {15723089}
}

@misc{Saengchote2022DecentralizedCompound,
    title = {{Decentralized lending and its users: Insights from Compound}},
    year = {2022},
    author = {Saengchote, Kanis},
    url = {https://etherscan.io/address/0xdac17f958d2ee523a2206206994597c13d831ec7#writeContract}
}

@misc{EuropeanParliament2014Directive2011/61/EU,
    title = {{Directive 2014/65/EU of the European Parliament and of the Council of 15 May 2014 on Markets in Financial Instruments and amending Directive 2002/92/EC and Directive 2011/61/EU}},
    year = {2014},
    booktitle = {Official Journal of the European Union},
    author = {{European Parliament} and of the European Union, Council},
    pages = {349--496},
    volume = {L 173},
    url = {https://eur-lex.europa.eu/legal-content/EN/TXT/?uri=CELEX:32014L0065}
}

@article{Alt2020ElectronicMarkets,
    title = {{Electronic Markets on blockchain markets}},
    year = {2020},
    journal = {Electronic Markets},
    author = {Alt, Rainer},
    month = {12},
    pages = {181--188},
    volume = {30},
    doi = {10.1007/s12525-020-00428-1}
}

@misc{Radomski2018ERC-1155:Standard,
    title = {{ERC-1155: Multi Token Standard}},
    year = {2018},
    author = {Radomski, Witek and {et al.}},
    howpublished = {https://eips.ethereum.org/EIPS/eip-1155}
}

@misc{VogelstellerERC-20:Standard,
    title = {{ERC-20: Token Standard}},
    author = {Vogelsteller, Fabian and Buterin, Vitalik},
    howpublished = {https://eips.ethereum.org/EIPS/eip-20}
}

@misc{Entriken2018ERC-721:Standard,
    title = {{ERC-721: Non-Fungible Token Standard}},
    year = {2018},
    author = {Entriken, William and Shirley, Dieter and Evans, Jacob and Sachs, Nastassia},
    howpublished = {https://eips.ethereum.org/EIPS/eip-721}
}

@techreport{ESMA2024ESMA75453128700-1323Instruments,
    title = {{ESMA75453128700-1323 Final Report on the Guidelines on the conditions and criteria for the qualification of crypto-assets as financial instruments}},
    year = {2024},
    author = {{ESMA}},
    month = {12},
    url = {www.esma.europa.eu},
    institution = {ESMA}
}

@techreport{U.S.SecuritiesandExchangeCommissionSEC2019FrameworkAssets,
    title = {{Framework for ``Investment Contract'' Analysis of Digital Assets}},
    year = {2019},
    author = {{U.S. Securities and Exchange Commission (SEC)}},
    month = {4},
    institution = {U.S. Securities and Exchange Commission}
}

@techreport{InterWorkAlliance2020InterWorkFinal,
    title = {{InterWork Alliance’s Token Taxonomy Fact Card (Final)}},
    year = {2020},
    author = {{InterWork Alliance}},
    institution = {InterWork Alliance}
}

@techreport{IOSCOIOSCOCommissions,
    title = {{IOSCO DECENTRALIZED FINANCE REPORT Public Report The Board of the International Organization of Securities Commissions}},
    author = {{IOSCO}},
    url = {https://www.iosco.org/library/pubdocs/pdf/IOSCOPD554.pdf.}
}

@article{Schuler2024OnFinance,
    title = {{On DeFi and On-Chain CeFi: How (Not) to Regulate Decentralized Finance}},
    year = {2024},
    journal = {Journal of Financial Regulation},
    author = {Schuler, Katrin and Cloots, Ann Sofie and Sch{\"{a}}r, Fabian},
    number = {2},
    month = {9},
    pages = {213--242},
    volume = {10},
    publisher = {Oxford University Press},
    doi = {10.1093/jfr/fjad014},
    issn = {20534841}
}

@inproceedings{Luo2025PiercingReappraised,
    title = {{Piercing the Veil of TVL: DeFi Reappraised}},
    year = {2025},
    booktitle = {Financial Cryptography and Data Security 2025},
    author = {Luo, Yichen and Feng, Yebo and Xu, Jiahua and Tasca, Paolo},
    month = {7},
    url = {http://arxiv.org/abs/2404.11745},
    arxivId = {2404.11745}
}

@article{Hartwich2024ProbablyTokens,
    title = {{Probably Something: A Multi-Layer Taxonomy of Non-Fungible Tokens}},
    year = {2024},
    journal = {Internet Research},
    author = {Hartwich, Felix and Begus, Samo and Thiebes, Scott and Sunyaev, Ali}
}

@inproceedings{Yuyama2024ProposalRegulation,
    title = {{Proposal of Principles of DeFi Disclosure and Regulation}},
    year = {2024},
    booktitle = {FC 2023 Workshops},
    author = {Yuyama, Tomonori and Katayama, Ken and Brigner, Paul},
    pages = {141--164},
    publisher = {International Financial Cryptography Association}
}

@article{Xu2023ReapProtocols,
    title = {{Reap the Harvest on Blockchain: A Survey of Yield Farming Protocols}},
    year = {2023},
    journal = {IEEE Transactions on Network and Service Management},
    author = {Xu, Jiahua and Feng, Yebo},
    number = {1},
    month = {3},
    pages = {858--869},
    volume = {20},
    publisher = {Institute of Electrical and Electronics Engineers Inc.},
    doi = {10.1109/TNSM.2022.3222815},
    issn = {19324537},
    arxivId = {2210.04194}
}

@misc{Odinet2025REGULATING2026,
    title = {{REGULATING STABLECOINS: COMPARING MICAR AND THE GENIUS ACT NOTRE DAME LAW REVIEW REFLECTION (forthcoming 2026)}},
    year = {2025},
    author = {Odinet, Christopher K and Tosato, Andrea},
    url = {https://www.whitehouse.gov/fact-sheets/2025/07/fact-sheet-president-}
}

@techreport{HoganLovells2025RegulatingGuide,
    title = {{Regulating Staking: A Comparative Guide}},
    year = {2025},
    author = {{Hogan Lovells}},
    url = {https://www.hoganlovells.com},
    institution = {Hogan Lovells International LLP}
}

@misc{EuropeanParliament2023Regulation2019/1937,
    title = {{Regulation (EU) 2023/1114 of the European Parliament and of the Council of 31 May 2023 on Markets in Crypto-assets, and amending Regulations (EU) No 1093/2010 and (EU) No 1095/2010 and Directives 2013/36/EU and (EU) 2019/1937}},
    year = {2023},
    booktitle = {Official Journal of the European Union},
    author = {{European Parliament} and of the European Union, Council},
    pages = {40--205},
    volume = {L 150},
    url = {https://eur-lex.europa.eu/legal-content/EN/TXT/?uri=CELEX:32023R1114}
}

@article{Liu2021RisksCryptocurrency,
    title = {{Risks and Returns of Cryptocurrency}},
    year = {2021},
    journal = {Journal of Finance},
    author = {Liu, Yukun and Tsyvinski, Aleh and Wu, Xi}
}

@misc{SupremeCourtoftheUnitedStates1946SECCo.,
    title = {{SEC v. W.J. Howey Co.}},
    year = {1946},
    author = {{Supreme Court of the United States}},
    pages = {293--299},
    volume = {328},
    url = {https://supreme.justia.com/cases/federal/us/328/293/}
}

@inproceedings{Ovezik2024SoK:Decentralization,
    title = {{SoK: A Stratified Approach to Blockchain Decentralization}},
    year = {2024},
    booktitle = {Financial Cryptography and Data Security},
    author = {Ovezik, Christina and Karakostas, Dimitris and Kiayias, Aggelos},
    month = {2},
    pages = {128},
    url = {http://arxiv.org/abs/2211.01291},
    arxivId = {2211.01291}
}

@article{Zhang2023SoK:Decentralization,
    title = {{SoK: Blockchain Decentralization}},
    year = {2023},
    journal = {arXiv},
    author = {Zhang, Luyao and Ma, Xinshi and Liu, Yulin},
    month = {8},
    url = {http://arxiv.org/abs/2205.04256},
    arxivId = {2205.04256}
}

@article{Xu2022SoK:Protocols,
    title = {{SoK: Decentralized Exchanges (DEX) with Automated Market Maker (AMM) Protocols}},
    year = {2022},
    journal = {ACM Computing Surveys},
    author = {Xu, Jiahua and Paruch, Krzysztof and Cousaert, Simon and Feng, Yebo},
    month = {11},
    url = {http://arxiv.org/abs/2103.12732 http://dx.doi.org/10.1145/3570639},
    doi = {10.1145/3570639},
    arxivId = {2103.12732}
}

@inproceedings{Werner2022SoK:DeFi,
    title = {{SoK: Decentralized Finance (DeFi)}},
    year = {2022},
    booktitle = {Proceedings of the 4th ACM Conference on Advances in Financial Technologies (AFT ’22)},
    author = {Werner, Sam M. and Perez, Daniel and Gudgeon, Lewis and Klages-Mundt, Ariah and Harz, Dominik and Knottenbelt, William J.},
    month = {9},
    pages = {30--46},
    publisher = {Association for Computing Machinery},
    url = {http://arxiv.org/abs/2101.08778},
    doi = {10.1145/3558535.3559780},
    arxivId = {2101.08778}
}

@unpublished{Gogol2024SoK:Risks,
    title = {{SoK: Decentralized Finance (DeFi) -- Fundamentals, Taxonomy and Risks}},
    year = {2024},
    author = {Gogol, Krzysztof and Killer, Christian and Schlosser, Malte and Bocek, Thomas and Stiller, Burkhard and Tessone, Claudio},
    month = {4},
    url = {http://arxiv.org/abs/2404.11281},
    arxivId = {2404.11281}
}

@inproceedings{Ovezik2025SoK:Decentralization,
    title = {{SoK: Measuring Blockchain Decentralization}},
    year = {2025},
    booktitle = {International Conference on Applied Cryptography and Network Security},
    author = {Ovezik, Christina and Karakostas, Dimitris and Milad, Mary and Kiayias, Aggelos and Woods, Daniel W.},
    month = {2},
    pages = {184},
    publisher = {Springer},
    url = {http://arxiv.org/abs/2501.18279},
    arxivId = {2501.18279}
}

@unpublished{Ling2025SoK:LEGO,
    title = {{SoK: Stablecoin Designs, Risks, and the Stablecoin LEGO}},
    year = {2025},
    author = {Ling, Shengchen and Du, Yuefeng and Zhou, Yajin and Wu, Lei and Wang, Cong and Jia, Xiaohua and Yan, Houmin},
    month = {6},
    url = {http://arxiv.org/abs/2506.17622},
    arxivId = {2506.17622}
}

@inproceedings{Cousaert2022SoK:DeFi,
    title = {{SoK: Yield Aggregators in DeFi}},
    year = {2022},
    booktitle = {IEEE International Conference on Blockchain and Cryptocurrency, ICBC 2022},
    author = {Cousaert, Simon and Xu, Jiahua and Matsui, Toshiko},
    publisher = {Institute of Electrical and Electronics Engineers Inc.},
    isbn = {9781665495387},
    doi = {10.1109/ICBC54727.2022.9805523},
    arxivId = {2105.13891}
}

@article{Sai2021TaxonomyReview,
    title = {{Taxonomy of centralization in public blockchain systems: A systematic literature review}},
    year = {2021},
    journal = {Information Processing and Management},
    author = {Sai, Ashish Rajendra and Buckley, Jim and Fitzgerald, Brian and Gear, Andrew Le},
    number = {4},
    month = {7},
    volume = {58},
    publisher = {Elsevier Ltd},
    doi = {10.1016/j.ipm.2021.102584},
    issn = {03064573},
    arxivId = {2009.12542}
}

@unpublished{Annunziata2024TAXONOMYANALYSIS,
    title = {{TAXONOMY OF CRYPTO-ASSETS AND FINANCIAL INSTRUMENTS: WHERE DO WE STAND ? SOME LESSONS FROM A USA-EU COMPARATIVE ANALYSIS}},
    year = {2024},
    booktitle = {Bocconi Legal Studies Research Paper},
    author = {Annunziata, Filippo},
    number = {No. 4989905},
    institution = {Bocconi Legal Studies Research Paper}
}

@techreport{Breydo2024TheElusive,
    title = {{The Broken Token Problem: Why Crypto Classification Remains The Broken Token Problem: Why Crypto Classification Remains Elusive Elusive}},
    year = {2024},
    author = {Breydo, Lev E},
    url = {https://scholarship.law.wm.edu/facpubs},
    institution = {William {\&} Mary Law School}
}

@article{Scharnowski2025TheDiscovery,
    title = {{The Economics of Liquid Staking Derivatives: Basis Determinants and Price Discovery}},
    year = {2025},
    journal = {Journal of Futures Markets},
    author = {Scharnowski, Stefan and Jahanshahloo, Hossein},
    number = {2},
    month = {2},
    pages = {91--117},
    volume = {45},
    publisher = {John Wiley and Sons Inc},
    doi = {10.1002/fut.22556},
    issn = {10969934}
}

@misc{Krause2025TheRegulation,
    title = {{The GENIUS Act: A New Era of U.S. Stablecoin Regulation}},
    year = {2025},
    author = {Krause, David},
    month = {12},
    doi = {10.13140/RG.2.2.12228.74884}
}

@article{Weking2019ThePatterns,
    title = {{The impact of blockchain technology on business models – a taxonomy and archetypal patterns}},
    year = {2019},
    journal = {Electronic Markets},
    author = {Weking, Jörg and Mandalenakis, Michael and Hein, Andreas and Hermes, Sebastian and B{\"{o}}hm, Markus and Krcmar, Helmut},
    number = {},
    month = {12},
    pages = {285--305},
    volume = {30},
    url = {https://doi.org/10.1007/s12525-019-00386-3},
    doi = {10.1007/s12525-019-00386-3/Published}
}

@techreport{Auer2023TheDeFi,
    title = {{The Technology of Decentralized Finance (DeFi)}},
    year = {2023},
    author = {Auer, Raphael and Haslhofer, Bernhard and Kitzler, Stefan and Saggese, Pietro and Victor, Friedhelm},
    url = {www.bis.org},
    institution = {BIS}
}

@misc{Peirce2025ThereHere,
    title = {{There Must Be Some Way Out of Here}},
    year = {2025},
    author = {Peirce, Hester M},
    url = {https://www.sec.gov/news/speech},
    howpublished = {Speech, U.S. Securities and Exchange Commission}
}

@article{Sumanov2025TokenManifestation,
    title = {{Token Classification Framework: Considering the Origins of Value and their Mechanisms of Manifestation}},
    year = {2025},
    journal = {The Journal of The British Blockchain Association},
    author = {Sumanov, Vasily and Desmarais, Jean-Luc},
    month = {12},
    pages = {1--10},
    volume = {8},
    doi = {10.31585/jbba-8-1-(3)2025}
}

@article{Nadler2023TornadoPolicymakers,
    title = {{Tornado Cash and Blockchain Privacy: A Primer for Economists and Policymakers}},
    year = {2023},
    journal = {Federal Reserve Bank of St. Louis Review},
    author = {Nadler, Matthias and Sch{\"{a}}r, Fabian},
    number = {2},
    month = {4},
    pages = {122--136},
    volume = {105},
    publisher = {Federal Reserve Bank of St.Louis},
    doi = {10.20955/r.105.122-136},
    issn = {00149187}
}

@article{Gimpel2018UnderstandingOfferings,
    title = {{Understanding FinTech start-ups – A taxonomy of consumer-oriented service offerings}},
    year = {2018},
    journal = {Electronic Markets},
    author = {Gimpel, Henner and Rau, Daniel and Roeglinger, Maximilian},
    month = {12},
    pages = {245--264},
    volume = {28},
    doi = {10.1007/s12525-017-0275-0}
}

@article{Tonnissen2020UnderstandingStart-ups,
    title = {{Understanding token-based ecosystems – a taxonomy of blockchain-based business models of start-ups}},
    year = {2020},
    journal = {Electronic Markets},
    author = {T{\"{o}}nnissen, Stefan and Beinke, Jan and Teuteberg, Frank},
    month = {12},
    volume = {30},
    doi = {10.1007/s12525-020-00396-6}
}

@article{Giulio2021WrappingTokens,
    title = {{Wrapping trust for interoperability. A study of wrapped tokens}},
    year = {2021},
    journal = {Information},
    author = {Giulio, Caldarelli},
    arxivId = {2109.06847}
}
